\pdfoutput=1
\documentclass[twocolumn]{aastex631}
\graphicspath{{./}{figures/}}
\usepackage[T1]{fontenc}

\usepackage{soul}
\usepackage{amsmath}

\shorttitle{Uncountable HFSs in the W33 complex}
\shortauthors{L.~K. Dewangan et al.}

\begin{document}
\title{Discovery of a rich population of compact hub-filament systems in a single star-forming complex}

\author[0000-0001-6725-0483]{L.~K. Dewangan}
\affiliation{Astronomy \& Astrophysics Division, Physical Research Laboratory, Navrangpura, Ahmedabad 380009, India.}

\author[0000-0002-6740-7425]{Ram~K. Yadav}
\affiliation{National Astronomical Research Institute of Thailand (Public Organization), 260 Moo 4, T. Donkaew, A. Maerim, Chiangmai 50180, Thailand.}
\email{ram$\_$kesh@narit.or.th;  lokeshd@prl.res.in}

\author[0000-0001-5731-3057]{Saurabh Sharma}
\affiliation{Aryabhatta Research Institute of Observational Sciences, Manora Peak, Nainital 263002, India.}

\author[0009-0001-2896-1896]{O.~R. Jadhav}
\affiliation{Astronomy \& Astrophysics Division, Physical Research Laboratory, Navrangpura, Ahmedabad 380009, India.}
\affiliation{Indian Institute of Technology Gandhinagar Palaj, Gandhinagar 382355, India.}

\author[0000-0002-7367-9355]{A.~K. Maity}
\affiliation{Astronomy \& Astrophysics Division, Physical Research Laboratory, Navrangpura, Ahmedabad 380009, India.}

\author[0000-0002-6622-8396]{Paul F. Goldsmith}
\affiliation{Jet Propulsion Laboratory, California Institute of Technology, 4800 Oak Grove Drive, Pasadena, CA 91109, USA.}

\author{G. Panchal}
\affiliation{75, Jogani Nagar Society, Adajan, Surat 395 009, Gujarat, India.}

\begin{abstract}
We report the discovery of 45 compact hub-filament systems (HFSs; median size $\sim$2.4 pc) in infrared-dark clouds (IRDCs) in the W33 complex, located at the junction of the Scutum and Norma spiral arms. Using {\it Spitzer} 8 and 24 micron, and unWISE 12  $\mu$m images, HFSs are identified as regions where three or more filaments converge onto a central hub, appearing as absorption features toward IRDCs. In each IRDC, HFSs mainly lie  at the intersections of elongated substructures, associated with groups of protostars and lacking radio continuum emission.  
Minimum Spanning Tree (MST) analysis shows that protostars are closely associated with the HFSs, with protostellar core separations of $\leq$ 0.7 pc, indicating strong clustering within fragmented structures. 
The HFSs form two main groupings spanning 10--15 pc, with member separations of 1--3.3 pc. Around 65\% are tightly clustered ($<$ 2 pc), exhibiting rich small-scale structures and emphasizing the uniqueness of the complex. MST analysis of ALMAGAL 1.38 mm continuum cores---predominantly low-mass and embedded in ten HFSs---reveals a median core separation of $\sim$0.03 pc.
The protostellar spacing ($\sim$0.7 pc) significantly exceeds the thermal Jeans length ($\sim$0.08 pc for temperature $\sim$18 K and density $\sim$10$^{5}$ cm$^{-3}$), whereas the core spacing is smaller than the Jeans length, suggesting that thermal fragmentation may influence core formation but alone cannot explain the larger-scale protostellar distribution. All these findings together support a picture in which fragments of clouds/filaments form clumps hosting compact HFSs that facilitate efficient and clustered star formation, often yielding massive stars. 
\end{abstract}
%
\section{Introduction} 
\label{sec:intro}
With the advent of multi-scale infrared, sub-millimeter, and millimeter observations, significant progress has 
been made in understanding the formation of massive OB stars ($>$ 8 M$_{\odot}$) and young stellar clusters. 
In particular, hub-filament systems \citep[HFSs;][]{myers09} have emerged as key structures characterized by a network of filaments converging toward a central hub---typically the birthplace of massive stars and young stellar clusters \citep{Motte+2018,kumar20,padoan20,semadeni09,semadeni17,semadeni19,zhou22,yang23,maity23,maity25}. 
In HFSs, gas flows channeled along the filaments, driven by gravity and/or turbulence, are thought to play an important role in accumulating mass 
in the central hub \citep{liu23,bhadari23,bhadari25}, where initial star-forming seeds form and grow into massive stars \citep{Motte+2018}. 
However, feedback from massive OB stars can reshape early hub-filament configurations and 
influence the initial conditions for massive star formation (MSF). 
Therefore, studying young HFSs, where stellar feedback is still limited, is essential for understanding the earliest stages of hub evolution and the onset of MSF \citep[e.g.,][]{dewangan24}.
Despite their importance, such early-phase HFSs remain largely underexplored in the literature. 

While earlier studies have primarily focused on single/isolated HFSs across a wide range of spatial scales \citep{schneider12,liu12,peretto13,peretto14,morales19,dewangan20,dewangan23gl,dewangan25s,wang22,xu23,zhou23x,bhadari25}, recent works have revealed the presence of two or more HFSs within individual star-forming complexes \citep[e.g.,][]{bhadari22,dewangan23,dewangan24,verma23,zhou23x,maity25x}. 
Despite the existing findings related to HFSs, the earlier studies leave some important open questions: How many HFSs can coexist within an individual star-forming site or filamentary cloud? Are these HFSs small-scale/compact, large-scale, or a combination of both? What is their collective role in the star formation process? Which physical mechanisms govern their formation? And does the presence of multiple HFSs enhance star formation activity? To explore these questions, the present study investigates the filamentary structures in the W33 complex. 

A large-scale system ($\sim$50 pc $\times$ 37 pc) encompassing the W33 complex has been reported using a multi-wavelength approach \citep{dewangan20w33}. They adopted a distance of $\sim$2.6 kpc for the W33 complex (see Section~\ref{s1sec:d} for more details), which is consistent within uncertainties with the previously reported trigonometric parallactic distance of $2.4^{+0.17}_{-0.15}$ kpc \citep{immer13}. 
A distance of 2.6 kpc to the W33 complex is adopted in this paper. The large-scale structure of the W33 region has been studied in the two lowest rotational transitions of CO by \citet{goldsmith83} who derived a total mass of 0.2--2.0 $\times$ 10$^{6}$ M$_{\odot}$. Within this system, two distinct environments have been identified, characterized by the presence of filamentary structures and H\,{\sc ii} regions \citep[e.g.,][]{dewangan20w33,liu21,zhou23}. Using the {\it Herschel} dust temperature ($T_\mathrm{dust}$) map, Figure~\ref{fig1}a shows prominent filamentary structures along with regions of warm dust emission associated with H\,{\sc ii} regions in the W33 complex, where previously reported massive dust clumps (W33 Main, W33A, W33B, W33 Main1, W33A1, and W33B1) are marked by filled stars \citep[from][]{immer14,kohno18}.  The positions of a large number of 870 $\mu$m dust clumps \citep{urquhart18} are shown in Figure~\ref{fig1}a, which were collected from the Atacama Pathfinder Experiment (APEX) Telescope Large Area Survey of the GALaxy \citep[ATLASGAL; resolution $\sim$19\rlap.{$''$}2;][]{schuller09} survey.  
 
Numerous Class~I protostars, with a mean age of $\sim$0.44 Myr \citep{evans09}, along with their clusterings, were reported toward filamentary 
structures and H\,{\sc ii} regions \citep{dewangan20w33,liu21}. 
This selected area around W33 is recognized for active MSF and the presence of young stellar clusters. 
A scenario involving converging/colliding flows from two distinct velocity components was proposed to explain the observed signatures of star formation activity in the system \citep[e.g.,][]{dewangan20w33}. Despite the existing studies in W33, no dedicated effort has yet been made to investigate the filamentary structures using high resolution infrared data, which is essential for probing the underlying star formation processes. 

In this study, we employ multi-wavelength, multi-scale observations to discover a rich population 
of compact HFSs toward dark filamentary structures in the W33 complex. 
These HFSs show clear signatures of MSF and active protostellar clustering. 
To our knowledge, this is the first report of such a high concentration of small-scale HFSs (extent $\sim$2--3 pc) 
within a single star-forming complex, providing valuable insights into the physical mechanisms driving intense star formation. 

Section~\ref{sec:obser} describes the observational data sets employed in this work, 
while Section~\ref{sec:data} presents the results obtained from the multi-wavelength analysis. 
A discussion of the implications of these findings concerning the W33 complex is given in Section~\ref{sec:disc}. 
Finally, Section~\ref{sec:conc} summarizes the key results and conclusion of the study.
\section{Data and analysis} 
\label{sec:obser}
We conducted an extensive multi-wavelength analysis using observational data spanning 
from near-infrared (NIR) to radio wavelengths. 
These datasets include: the {\it Spitzer} Galactic Legacy Infrared Mid-Plane Survey Extraordinaire \citep[GLIMPSE; resolution $\sim$2$''$;][]{benjamin03} at 3.6--8.0 $\mu$m, 
the Unblurred Coadds of the Wide-field Infrared Survey Explorer Imaging (unWISE) survey \citep[resolution $\sim$6$''$;][]{Lang_2014} at 12 $\mu$m, the {\it Spitzer} MIPS Inner Galactic Plane Survey \citep[MIPSGAL; resolution $\sim$6$''$;][]{carey05} at 24 $\mu$m, the South African Radio Astronomy Observatory (SARAO) MeerKAT Galactic Plane Survey \citep[SMGPS; resolution $\sim$8$''$;][]{Goedhart_2024} at 1.3 GHz, the NRAO VLA Sky Survey \citep[NVSS; resolution $\sim$45$''$;][]{NVSS} at 1.4 GHz, and Radio Ammonia Mid-Plane Survey \citep[RAMPS; resolution $\sim$34\rlap.{$''$}7;][]{hogge18} for NH$_{3}$(1,1). 
We also utilized the publicly available {\it Herschel} $T_\mathrm{dust}$ map (resolution $\sim$12$''$), generated using the Bayesian {\it PPMAP} technique \citep{marsh15,marsh17}. 
\section{Results}
\label{sec:data}
\subsection{Identification of compact HFSs in W33}
\label{s1sec:d} 
\begin{figure*}
\center
\includegraphics[width=0.5\linewidth]{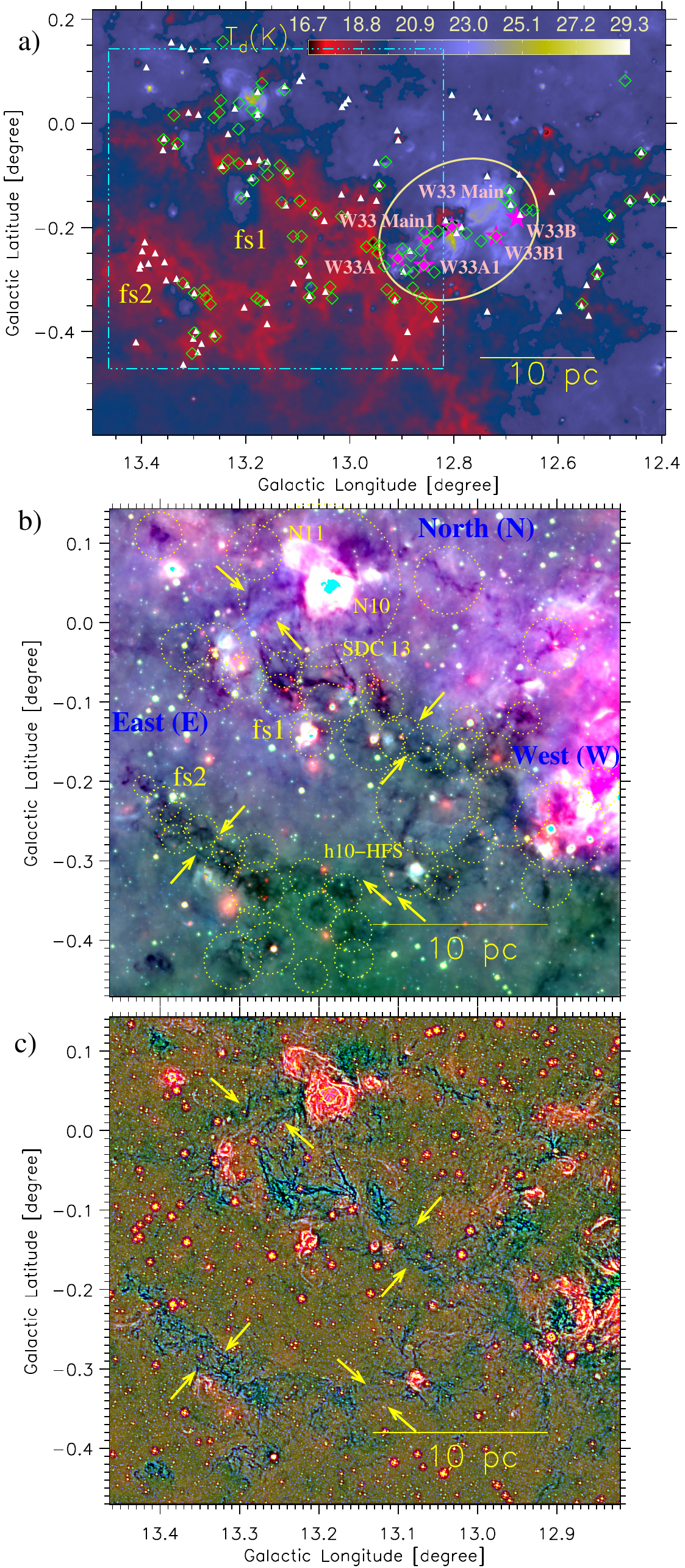}
\caption{a) The panel shows the {\it Herschel} 
temperature map of a larger area ($\sim$1$^\circ$.1 $\times$ 0$^\circ$.815 ($\sim$50 pc $\times$ 37 pc 
at  {\it d} $\sim$2.6 kpc), centered at Galactic coordinates {\it l} = 12$^\circ$.946, {\it b} = $-$0$^\circ$.1914. The area includes the W33 complex (indicated by the ellipse). The map is overlaid with the positions of the IRDCs (\citealt{peretto16}; see white filled triangles), massive clumps (magenta filled stars representing  W33 Main, W33A, W33B, W33 Main1, W33A1, and W33B1; \citealt{immer14}), and 870 $\mu$m dust continuum clumps (green diamonds; \citealt{urquhart18}). The area hosting elongated filaments, fs1 and fs2, is outlined with the dot-dashed box. b) The panel displays a three-color composite map (red: 24 $\mu$m, green: 12 $\mu$m, blue: 8 $\mu$m) of the area highlighted by the dot-dashed box in Figure~\ref{fig1}a. Dotted circles mark the positions of HFSs identified in the MIR images (see Table~\ref{tab1}).
Two sub-structures in absorption are indicated by arrows. c) The panel presents a three-color composite map (red: 12 $\mu$m, green: DoG processed 12 $\mu$m, blue: DoG processed 8 $\mu$m). The scale bar corresponding to 10 pc at {\it d} $\sim$2.6 kpc is shown in each panel.}
\label{fig1}
\end{figure*}

Figure~\ref{fig1}a displays the {\it Herschel} $T_\mathrm{dust}$ map overlaid with the positions of the infrared-dark clouds (IRDCs; white filled triangles; \citealt{peretto16}) and the ATLASGAL clumps (diamonds; \citealt{urquhart18}). \citet{urquhart18} resolved the kinematic distance ambiguity 
of the ATLASGAL clumps by combining H\,{\sc i} absorption analysis, maser parallax measurements, and spectroscopic data (see their paper for further details). In addition, they employed a friends-of-friends clustering algorithm to examine the large-scale organization of clumps. This method associates clumps that are close in Galactic longitude, latitude, and line-of-sight velocity (i.e., within l--b--v space), thereby identifying groups that likely represent coherent star-forming complexes. 
Using this approach, they identified a cluster of clumps associated with the W33 complex 
(see Figures B4 and C1 in \citealt{urquhart18}), located at a distance of $\sim$2.6 kpc and exhibiting a velocity range of $\sim$30--56 km s$^{-1}$ \citep[see also][]{dewangan20w33}. 

IRDCs, composed of dense, cold gas and dust that obscure background infrared emission, are ideal for probing the initial conditions of MSF \citep[e.g.,][]{ragan09}. 
The dot-dashed box in Figure~\ref{fig1}a highlights an area dominated by cold dust emission ($T_{\rm d}$ $\sim$17--19~K), containing numerous dust clumps and IRDCs, while warmer dust emission ($T_{\rm d}$ $\sim$20--30~K) appears near known H\,{\sc ii} regions (ellipse and filled symbols).
A mid-infrared (MIR) three-color composite map (red: 24 $\mu$m, green: 12 $\mu$m, blue: 8 $\mu$m; resolution $\sim$2$''$--6$''$) is presented in Figure~\ref{fig1}b, displaying prominent dark filamentary structures (fs1 and fs2; $T_{\rm d}$ $\sim$17--19~K; see Figure~\ref{fig1}a) alongside bright, extended MIR-emitting regions. 
Some previously reported sources (e.g., N10 and N11 \citep{dewangan23gl}; SDC~13 \citep{wang22}) are also indicated in Figure~\ref{fig1}b. 
It is noted that Figure~\ref{fig1}b is presented only toward the area hosting filamentary structures, where regions fs1, fs2, North (N), West (W), and East (E) are also marked. 

The MIR images provide a detailed view of the embedded structures within the IRDCs.
Through visual inspection, we identify 49 systems in which three or more filaments converge toward a central area/hub. The filaments are seen as cold, small-scale, high-aspect-ratio features in MIR absorption, whereas the hub appears more compact (or round) with a relatively low aspect ratio. 
Such a configuration--commonly described as the junction of three or more filaments \citep[e.g.,][]{kumar20, zhou24}--is the defining characteristic of HFSs. Accordingly, we classify these systems as HFSs, marking their locations with dotted circles in Figure~\ref{fig1}b. The diameter of each circle represents the size of the corresponding HFS. The clear morphological contrast between the compact hubs and the elongated filaments serves as a potential condition for distinguishing genuine HFSs from unrelated dark features. The positions and sizes of the identified HFSs are summarized in Table~\ref{tab1}. The spatial extents or sizes of these HFSs, including both their central hubs and converging filaments, range between 1.1 and 9.3 pc, with median and mean values of 2.4 pc and 2.8 pc, respectively. These measurements indicate that the identified systems are compact in nature. 
Among 49 HFSs, we have found 24, 18, 5, and 2 HFSs toward fs1, fs2, North, and West regions, respectively. Their size ranges are 1.1--5.9 pc for fs1 (median: 2.6 pc; average: 2.8 pc), 1.2--3.3 pc for fs2 (median and average: 2.1 pc), 2.7--9.3 pc for the North (median: 3.2 pc; average: 4.4 pc), and 2.3--5.1 pc for the West (median: 5.1 pc; average: 3.7 pc).  
However, a few of the 49 HFSs, mainly those located in the 
northern direction (except h5; see Table~\ref{tab1}), may not be located at a distance of $\sim$2.6 kpc (see also Section~\ref{s2sec:dx}), 
where the ATLASGAL clumps are not present. Consequently, this study considers 45 HFSs embedded within IRDCs in a single star-forming complex.

Zoomed-in views of these HFSs are shown in Figure~\ref{fig:apx1gg} using the MIR 8 $\mu$m emission.  
The area  shown for each HFS refers to its respective size as listed in Table~\ref{tab1}. 
In many cases, the filaments and the hubs (or central compact regions) appear as dark features in all the MIR images due to absorption of background radiation, suggesting that the central hub is not yet significantly heated or illuminated by embedded powering stars. 
These systems exhibit similar morphological features across all MIR 
images (not shown). Likewise, at least four similar HFSs have been identified in the Galactic `Snake' IRDC G11.11-0.12 using the {\it Spitzer} 8 $\mu$m image \citep[see Figure~2 in][]{dewangan24}. In nearly half of the sample, the central hubs are associated with at least one point-like source that appears more prominent in the 24 $\mu$m image (see Figure~\ref{fig1}b), while the remaining hubs do not exhibit any 24 $\mu$m source. 

Additionally, these MIR images reveal substructures, seen in absorption, toward both the filaments (fs1 and fs2; see arrows in Figure~\ref{fig1}b). To further probe the substructures, the 8 and 12 $\mu$m images were processed using the {\it Edge-DoG} algorithm, 
which applies the Difference of Gaussians (DoG) filtering technique \citep[e.g.,][]{Assirati2014}. 
The resulting processed images were used to produce the color composite map (red: 12 $\mu$m, 
green: DoG processed 12 $\mu$m, blue: DoG processed 8 $\mu$m) shown in Figure~\ref{fig1}c, 
where intertwined, elongated structures are evident (see arrows in Figure~\ref{fig1}c). Most of the HFSs appear to be situated at the intersection regions of the two elongated subfilamentary structures.  

\begin{figure*}
\center
\includegraphics[width=\textwidth]{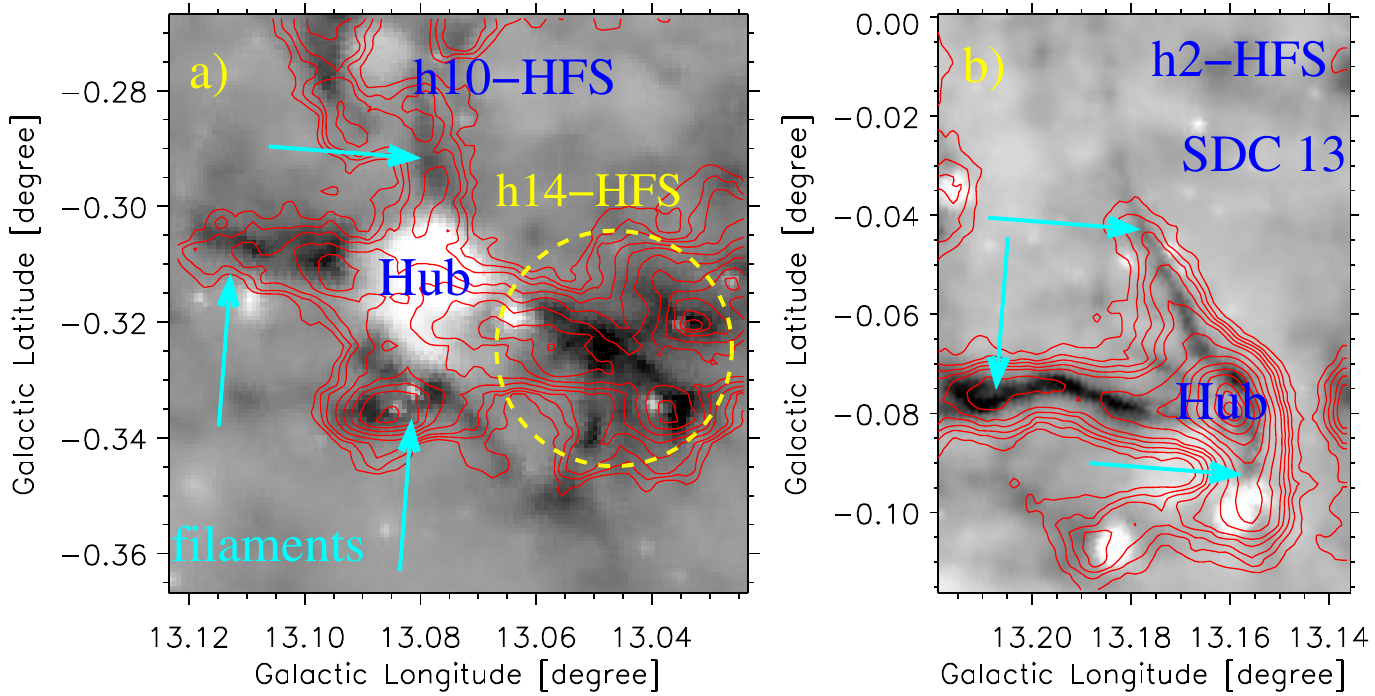}
\caption{a) 12 $\mu$m image of h10-HFS (including h14-HFS) overlaid with the NH$_{3}$ emission integrated over [36, 40] km s$^{-1}$. Contours are 3.706 K km s$^{-1}$ $\times$ [0.2, 0.25, 0.3, 0.4, 0.5, 0.6, 0.7, 0.8, 0.9, 0.95, 0.98]. b) 12 $\mu$m image of h2-HFS/SDC~13 overlaid with the NH$_{3}$ emission integrated over [35.3, 39.5] km s$^{-1}$. Contours are 6.542 K km s$^{-1}$ $\times$ [0.11, 0.16, 0.2, 0.25, 0.3, 0.4, 0.5, 0.6, 0.7, 0.8, 0.9, 0.95, 0.98].}
\label{fig1zx}
\end{figure*}

Figure~\ref{fig1zx}a presents the 12 $\mu$m image of h10-HFS (also including h14-HFS), overlaid 
with RAMPS NH$_{3}$(1,1) integrated emission contours in the velocity range of [36, 40] km s$^{-1}$. 
Similarly, Figure~\ref{fig1zx}b shows the 12 $\mu$m image of h2-HFS/SDC~13, with overlaid NH$_{3}$(1,1) integrated emission contours spanning [35.3, 39.5] km s$^{-1}$. 
In Figure~\ref{fig1zx}a, the molecular structure toward h14-HFS is not resolved. Nevertheless, both h10-HFS and h2-HFS reveal velocity-coherent NH$_{3}$ structures. 
It is important to note that the NH$_{3}$(1,1) data from the RAMPS survey are limited by relatively coarse beam size. Consequently, a detailed characterization of the gas distribution toward most of the HFSs reported in this work is not feasible (see Section~\ref{sec:clump} for further discussion).  
\subsection{Distribution of radio continuum emission and Hi-GAL clumps in W33}
\label{s2sec:dx}
Figure~\ref{fig:apx2}a presents the SMGPS 1.3 GHz continuum emission map overlaid with the NVSS 1.4 GHz continuum contour and the positions of HFSs. The positions of 870 $\mu$m clumps located at {\it d} $\sim$2.6 kpc \citep[from][]{urquhart18} are also indicated in Figure~\ref{fig:apx2}a.   
Note that the SMGPS radio continuum map (rms sensitivity $\sim$3--10 $\mu$Jy beam$^{-1}$; 
resolution $\sim$8$''$; \citealt{Goedhart_2024}) has better sensitivity and resolution compared to the NVSS radio continuum map (rms sensitivity $\sim$0.45 mJy beam$^{-1}$; resolution $\sim$45$''$; \citealt{NVSS}). Figure~\ref{fig:apx2}a shows that most of the selected HFSs lack associated radio continuum emission.   

The positions of the Hi-GAL clumps \citep[from][]{elia17} are examined with respect to the {\it Spitzer} 24 $\mu$m image (see Figure~\ref{fig:apx2}b). The positions of the selected HFSs are also marked in Figure~\ref{fig:apx2}b. 
Based on the absence of ATLASGAL clumps, it is likely that  some HFSs in the 
North region, excluding bubble N10, may not be situated at {\it d} $\sim$2.6~kpc (Figures~\ref{fig:apx2}a and~\ref{fig:apx2}b). 
In Figure~\ref{fig:apx2}b, the blue, yellow, green, cyan, and magenta squares show the positions of Hi-GAL clumps located toward regions fs1, fs2, North, West, and East, respectively. Various physical parameters of these clumps--such as mass, $T_\mathrm{dust}$, bolometric luminosity, and bolometric temperature--were previously derived by \citet{elia17}. 

The mass ranges of the clumps associated with the fs1 (blue squares), fs2 (yellow squares), North (green squares), West (cyan squares), and East (magenta squares) regions are 7.7--121.9 M$_{\odot}$, 5.5--341.0 M$_{\odot}$, 1.6--341.0 M$_{\odot}$, 3.1--151.6 M$_{\odot}$, and 1.7--71.6 M$_{\odot}$, respectively. The corresponding bolometric luminosity ranges are 2.1--1959.0 L$_{\odot}$ (fs1), 2.3--233.6 L$_{\odot}$ (fs2), 2.1--4825.7 L$_{\odot}$ (North), 2.7--4825.7 L$_{\odot}$ (West), and 4.6--22.4 L$_{\odot}$ (East). 
Similarly, the bolometric temperature ranges for fs1, fs2, North, West, and East are 13.6--56.7~K, 13.4--47.0~K, 13.4--77.9~K, 15.2--56.3~K, and 16.1--77.9~K, respectively. These results suggest that fs2 and fs1 likely represent a mix of early-stage clumps, with fs2 skewed toward less evolved, cooler sources. On the other hand, higher luminosity and temperature values are observed in Hi-GAL clumps associated with the North and West regions. 

To study HFSs in the Milky Way, \citet{kumar20} analyzed about 35000 Hi-GAL clumps from the catalog 
compiled by \citet{elia17}. Using {\it Herschel} 250 $\mu$m images (resolution $\sim$18$''$), they identified filamentary structures associated with these clumps, leading to the identification of around 3700 candidate HFSs. The filaments comprising these systems generally span lengths of 10--20 pc. 
Notably, all clumps with luminosities exceeding 10$^{4}$ L$_{\odot}$ and 10$^{5}$ L$_{\odot}$, at distances within 2 kpc and 5 kpc, 
respectively, were found to reside within the hubs of HFSs. 
These hubs commonly displayed a convergence of 3 to 7 filaments. 

In the direction of the selected target area, Figure~\ref{fig:apx2}b also displays the positions of candidate HFSs identified by \citet{kumar20} (see open and filled hexagons). 
The open hexagons represent sources located at {\it d} $\sim$2.6 kpc, while the filled hexagons correspond to sources at different distances (i.e., 1.9 kpc, 2.2 kpc, 4.8 kpc, and 14.6 kpc). From their catalog, we identify four candidate HFSs in fs1 and one in fs2, although the well-known HFS SDC~13 is notably missing. In this study, HFSs are identified using MIR images with angular resolutions of 2$''$--6$''$, allowing us to detect small-scale structures unresolved in the lower-resolution data ($\sim$18$''$) used by \citet{kumar20}.  Consequently, several HFSs selected in this paper are absent from their catalog. Differences in resolution, sensitivity, and wavelength coverage further explain why only a few HFSs from their catalog are recovered here.
\subsection{Spatial distributions of protostars and HFSs}
\label{sec:msta}
In recent years, Minimal Spanning Tree \citep[MST;][]{2009ApJS..184...18G} approach has been utilized to study the spatial distribution of embedded sources in star-forming regions \citep[e.g.,][]{2024AJ....167..106S,2017MNRAS.467.2943S,2016AJ....151..126S}.
\subsubsection{Minimum Spanning Tree analysis of protostars}
\label{sec:msta}
In this section, {\it Spitzer} photometric data at 3.6--24 $\mu$m of point-like sources were employed to study the embedded populations. 
Photometric magnitudes of point-like sources at {\it Spitzer} 3.6, 4.5, 5.8, and 24 $\mu$m bands were obtained from the GLIMPSE-I Spring '07 highly reliable catalog and the publicly available MIPSGAL catalog \citep[e.g.,][]{gutermuth15}, respectively. Embedded protostars (i.e., Class~I~sources and flat-spectrum~sources) were identified using the [3.6] $-$ [24]/[3.6] color-magnitude diagram \citep[e.g.,][]{guieu10,rebull11} and the [4.5]$-$[5.8] vs [3.6]$-$[4.5] color-color diagram \citep[e.g.,][]{hartmann05,getman07}.
Although these diagrams are not presented here, their methodology is described in \citet{dewangan15}. Figure~\ref{fig3}a overlays of the positions of the selected protostars (open circles) on the 12 $\mu$m image, which are seen toward filamentary structures and previously known H\,{\sc ii} regions. 

Figure~\ref{fig3}b displays the MST constructed to study the spatial distribution of protostars in the selected target area. 
Filled circles (in orange and blue) represent the positions of protostars, while the lines indicate the MST branches connecting them.
In Figure~\ref{fig3}b, the spatial distribution of protostars is highly inhomogeneous, featuring one prominent 
concentration in the central W33 region along with several smaller groupings evident across the region. 
To identify these substructures, we applied a surface density threshold determined by a critical branch length. 
The histogram of MST branch lengths (not shown) revealed distinct subregions embedded within a 
more diffuse stellar distribution, indicating areas of elevated local surface density. 

Based on this analysis, a threshold MST branch length of 0.7~pc was adopted, effectively encompassing the 
majority of branches and enabling the identification of protostars separated by less than this distance. These sources are interpreted as distinct star-forming surface density enhancements, and are marked with blue dots in Figure~\ref{fig3}b. We define groupings of at least four protostars within this threshold as ``core regions''. 
This approach has enabled us to effectively identify distinct protostellar cores within the target area, 
separating them from the surrounding diffuse protostar population. The most prominent groupings are located in the central W33 region, while numerous smaller cores are distributed across the field--many of which are spatially coincident with IRDCs. 
Notably, the protostars within these cores are very young, with a mean age of less than 1 Myr \citep{evans09}.
In the direction of IRDCs, the positions of HFSs, protostars (including ``core regions''), and dust clumps are presented in Figure~\ref{fig3}c, enabling an assessment of their spatial association.
\subsubsection{Minimum Spanning Tree of HFSs}
\label{suvsec:mstx}
To further study the properties of the identified HFSs and their spatial associations, the MST is constructed using their positions. Figure~\ref{mst}a presents the resulting MST of the selected HFSs. 
Solid lines connecting red and yellow dots represent MST branches 
that are shorter than the critical lengths of 2 pc for cores and 3.3 pc for active regions (ARs), respectively. Purple dots denote isolated HFSs that are not connected by branches within these critical thresholds. 
These HFSs are located in the North region. 

The critical lengths for the cores and ARs were derived based on the distribution of MST branch lengths. In Figure~\ref{mst}b, we show the histogram of the MST branch lengths for the HFSs. The distribution clearly shows two distinct peaks at 2 pc and 3.3 pc. 
Figure~\ref{mst}c presents the cumulative distribution of MST branch lengths, revealing noticeable changes in slope at 2 pc and 3.3 pc. This further reinforces these values as critical branch lengths. Accordingly, the same thresholds are applied to identify groupings in Figure~\ref{mst}a.

Two distinct groups of HFSs are identified in filamentary clouds extending approximately 10--15 pc in length (see Figure~\ref{mst}a). All HFSs in these groupings appear to be associated with larger star-forming complexes or embedded within a molecular cloud. HFSs not included in these clusters are regarded as isolated cases.
Separations between HFSs within these structures range from 1 to 3.3 pc. Approximately 65\% of them form compact clusters with separations below 2 pc, displaying a rich distribution of small-scale HFSs in the target area 
(see Figures~\ref{mst}b and~\ref{mst}c). However, a fraction of HFSs are located outside these clusters, possibly representing independently formed systems or different fragmentation stages.

MST analysis of protostars in the selected region reveals the presence of protostellar cores embedded within the HFSs, indicating a strong spatial association between protostars and these structures. These protostellar cores exhibit separations of 0.7 pc or less, highlighting a high degree of clustering within the fragmented structures.
\subsection{RAMPS NH$_{3}$(1,1) emission toward W33}
\label{sec:clump}
The RAMPS NH$_{3}$(1,1) line data are examined in a velocity range of [31.9, 43.5] km s$^{-1}$ toward the W33 complex. In the direction of the selected target area, Figures~\ref{fig4}a and~\ref{fig4}b present the NH$_{3}$(1,1) moment-0 and moment-1 maps, respectively. The positions of HFSs and dust clumps are also marked in Figure~\ref{fig4}a. In general, NH$_{3}$ traces colder, chemically mature dense gas interiors better than CO isotopologues. The NH$_{3}$ emission is depicted toward both the IRDCs (including HFSs) and the central W33 region. 
From this result and the molecular structures in Figures~\ref{fig1zx}a and~\ref{fig1zx}b, we anticipate that the selected HFSs in our study are also likely to exhibit velocity-coherent structures. It suggests that the filaments converging toward the hubs are not merely chance alignments along the line of sight, but are physically connected components of a common system. 
To confirm this, high-resolution observations using dense gas tracers will be necessary. 

In the NH$_{3}$(1,1) moment-1 map, velocity variations are evident toward both the filaments, fs1 and fs2. We have also examined the moment-0 maps and position-velocity diagrams of NH$_{3}$(1,1) emission toward both the filaments separately (see Figures~\ref{fig:apx4}a--\ref{fig:apx4}d and also Appendix~\ref{susec:gsp}). The gas associated with fs1 is relatively more blueshifted compared to that associated with fs2, indicating a kinematic distinction between these filamentary structures. 
Both filaments contain several molecular clumps, where velocity spreads are evident (see Figures~\ref{fig:apx4}b and~\ref{fig:apx4}d). 

We also explored the $N$(H$_2$) map of the target area, which was derived using the existing $N$(NH$_3$) map collected from the RAMPS survey \citep[see][for more details]{hogge18}. Molecular clumps were identified using the $N$(H$_2$) map with the {\it astrodendro} \citep{Rosolowsky_2008ApJ} (see Figure~\ref{fig5}a and also Appendix~\ref{susec:gsp} for more details). 
The physical properties of these clumps--such as mass ($M_{\mathrm{clump}}$) and effective radius ($R_{\mathrm{eff}}$)--were determined and are listed in Table~\ref{tab2} (see also Appendix~\ref{susec:gsp}). By placing the clumps in the $M_{\mathrm{clump}}$--$R_{\mathrm{eff}}$ space and applying the Kauffmann \& Pillai (KP-10) condition for MSF \citep[][]{Kauffmann2010}, we find that all identified clumps, which are also distributed toward fs1 and fs2, satisfy the threshold required to potentially form massive stars (see Figure~\ref{fig5}b).

Using these RAMPS NH$_{3}$(1,1) moment-0 and moment-1 maps, along with associated data products (i.e., rotational temperature ($T_\mathrm{rot}$) and kinematic temperature ($T_\mathrm{kin}$) maps) and the {\it Herschel} $T_\mathrm{dust}$ map, we study the physical parameters for all 
identified HFSs. 
We calculated the average values of integrated NH$_{3}$(1,1) intensity, NH$_{3}$(1,1) velocity, 
$T_\mathrm{rot}$, $T_\mathrm{kin}$, and $T_\mathrm{dust}$ for all identified HFSs, as summarized in Table~\ref{tab1}. The average intensity and velocity were derived using the moment-0 and moment-1 maps, 
respectively (see Figures~\ref{fig3}a and~\ref{fig3}b). 
The average $T_\mathrm{dust}$ was estimated from the {\it Herschel} $T_\mathrm{dust}$ map (see Figure~\ref{fig1}a), while the existing $T_\mathrm{rot}$ and $T_\mathrm{kin}$ maps 
were used to estimate the corresponding average temperatures. The physical properties of these HFSs vary across regions. The velocity ranges are 35.1--38.4 km s$^{-1}$ for fs1, 38.1--40.2 km s$^{-1}$ for fs2, 33.3--38.5 km s$^{-1}$ for the North, and 36.4--36.5 km s$^{-1}$ for the West. $T_\mathrm{rot}$ values span 12.2--18.0 K (fs1), 12.4--13.2 K (fs2), 13.5--17.3 K (North), and 15.1--19.1 K (West), while $T_\mathrm{kin}$ values range from 12.9--21.2 K (fs1), 13.1--14.2 K (fs2), 14.6--20.1 K (North), and 16.8--23.7 K (West). $T_\mathrm{dust}$ values are found to lie between 18.3--20.3 K (fs1), 18.3--18.8 K (fs2), 20.0--20.4 K (North), and 20.1--20.5 K (West). 

The fs1 and fs2 regions, which contain the majority of HFSs, are characterized by lower $T_\mathrm{rot}$ and $T_\mathrm{dust}$, along with relatively compact structures. The gas in these regions is still cold and dense, and feedback effects from massive stars located in the W33 complex are minimal.  
\begin{table*}
\scriptsize
\centering
\caption{Positions of HFSs identified in the MIR images (see dotted circles in Figure~\ref{fig1}b). 
Table contains ID, positions, size/extent, averaged integrated NH$_{3}$(1,1) intensity ($I_\mathrm{NH3}$),  
averaged NH$_{3}$(1,1) velocity ($V_\mathrm{NH3}$), 
averaged rotational temperature ($T_\mathrm{rot}$), averaged kinematical temperature ($T_\mathrm{kin}$), and averaged dust temperature ($T_\mathrm{dust}$). 
Last column hosts names of regions (north (N), west (W), fs1, and fs2) where the HFSs are spatially associated (see Figure~\ref{fig1}b).}
\label{tab1}
\begin{tabular}{lcccccccccccccc}
\toprule
ID & $l$ &  $b$     &size      &$I_\mathrm{NH3}$   &     Velocity      &    $T_\mathrm{rot}$ &    $T_\mathrm{kin}$& $T_\mathrm{dust}$   &    Association                  \\
& (degree) &  (degree)  &(pc) & (K km s$^{-1}$)  &    (km s$^{-1}$)  &     (K)        &   (K)              &    (K) &               \\
\hline
  h1 &13.121 &  $-$0.100 & 2.47 &2.36  & 36.59   & 13.78 & 14.96& 18.57 &fs1   \\
  h2 &13.179 &  $-$0.076 & 4.40 &1.33  & 36.92   & 14.16 & 15.51& 19.15 &fs1   \\
  h3 &13.244 &  $-$0.059 & 2.05 &2.52  & 38.02   & 13.31 & 14.34& 18.59 &fs1   \\
  h4 &12.908 &  $-$0.029 & 3.11 &0.30  & 35.97   & 17.34 & 20.14& 20.36 &N   \\
  h5 &13.200 &   0.047   & 9.30 &0.40  & 38.47   & 16.25 & 18.59& 20.17 &N  \\
  h6 &13.333 &  $-$0.037 & 3.33 &0.49  & 35.43   & 14.74 & 16.29& 19.45 &fs1   \\
  h7 &13.336 &  $-$0.078 & 3.11 &0.50  & 36.16   & 13.95 & 15.21& 19.24 &fs1   \\
  h8 &13.249 &  $-$0.092 & 2.40 &2.88  & 37.01   & 13.80 & 15.01& 18.73 &fs1   \\
  h9 &12.942 &  $-$0.120 & 1.84 &1.91  & 36.64   & 12.22 & 12.94& 19.33 &fs1   \\
 h10 &13.085 &  $-$0.310 & 3.15 &1.23  & 38.14   & 15.01 & 16.69& 18.61 &fs1   \\
 h11 &13.131 &  $-$0.148 & 3.44 &0.91  & 37.63   & 15.01 & 16.68& 18.54 &fs1   \\
 h12 &12.896 &  $-$0.257 & 5.13 &6.16  & 36.36   & 15.08 & 16.78& 20.12 &W   \\
 h13 &12.848 &  $-$0.209 & 2.33 &4.94  & 36.53   & 19.15 & 23.75& 20.52 &W  \\
 h14 &13.045 &  $-$0.327 & 1.93 &2.44  & 38.38   & 13.08 & 14.04& 18.29 &fs1   \\
 h15 &13.169 &  $-$0.339 & 2.07 &1.88  & 38.95   & 12.47 & 13.26& 18.31 &fs2   \\
 h16 &13.277 &  $-$0.335 & 3.01 &4.05  & 39.84   & 12.78 & 13.65& 18.46 &fs2   \\
 h17 &13.159 &  $-$0.388 & 2.35 & --   &  --	 & --    &   -- & 18.40 &fs2  \\
 h18 &13.020 &  $-$0.165 & 2.96 &1.72  & 35.95   & 13.80 & 14.97& 18.49 &fs1  \\
 h19 &12.960 &  $-$0.231 & 4.56 &3.72  & 35.68   & 13.88 & 15.09& 18.60 &fs1  \\
 h20 &13.368 &  $-$0.031 & 2.75 &0.33  & 35.44   & 13.83 & 15.02& 18.97 &fs1  \\
 h21 &13.282 &   0.089   & 3.16 &0.06  & 34.17   & 13.89 & 15.11& 20.05 &N  \\
 h22 &13.006 &  $-$0.273 & 2.64 &1.72  & 35.07   & 12.53 & 13.35& 18.40 &fs1  \\
 h23 &13.380 &  $-$0.245 & 1.78 &2.23  & 38.79   & 12.36 & 13.13& 18.27 &fs2  \\
 h24 &13.349 &  $-$0.263 & 1.88 &2.05  & 39.28   & 12.63 & 13.46& 18.44 &fs2  \\
 h25 &13.322 &  $-$0.309 & 1.95 &3.32  & 39.41   & 12.85 & 13.75& 18.65 &fs2  \\
 h26 &13.380 &  $-$0.273 & 1.62 &0.81  & 39.45   & 13.21 & 14.19& 18.38 &fs2  \\
 h27 &13.396 &  $-$0.228 & 1.95 &1.47  & 38.93   & 12.48 & 13.27& 18.46 &fs2  \\
 h28 &13.037 &   0.053   & 3.85 &0.21  & 34.64   &  --   &   -- & 19.99 &N \\
 h29 &13.137 &  $-$0.077 & 1.07 &1.90  & 37.23   & 15.41 & 17.24& 18.86 &fs1  \\
 h30 &13.189 &  $-$0.107 & 2.75 &1.12  & 36.33   & 15.28 & 17.14& 19.39 &fs1  \\
 h31 &13.097 &  $-$0.146 & 2.61 &1.28  & 38.08   & 14.23 & 15.63& 18.35 &fs1  \\
 h32 &13.282 &  $-$0.379 & 1.93 &0.88  & 39.87   &  --   &   -- & 18.70 &fs2  \\
 h33 &13.309 &  $-$0.425 & 3.26 & --   & --	 &  --   &   -- & 18.55 &fs2  \\
 h34 &13.314 &  $-$0.287 & 1.88 &4.11  & 39.49   & 12.57 & 13.38& 18.53 &fs2  \\
 h35 &13.264 &  $-$0.410 & 2.94 & --   &  --	 & --    &   -- & 18.43 &fs2  \\
 h36 &13.064 &  $-$0.158 & 2.57 &1.32  & 36.80   & 12.79 & 13.67& 18.33 &fs1  \\
 h37 &12.981 &  $-$0.163 & 2.23 &2.67  & 35.69   & 13.83 & 15.01& 18.45 &fs1  \\
 h38 &13.063 &  $-$0.227 & 5.89 &0.85  & 36.84   & 13.19 & 14.18& 18.58 &fs1  \\
 h39 &13.402 &   0.110   & 2.68 &0.12  & 33.34   & 13.53 & 14.63& 20.35 &N  \\
 h40 &13.276 &  $-$0.290 & 2.18 &1.23  & 40.25   & 13.00 & 13.97& 18.73 &fs2  \\
 h41 &13.156 &  $-$0.423 & 2.21 & --   &  --	 & --    &   -- & 18.51 &fs2  \\
 h42 &12.913 &  $-$0.328 & 2.84 &2.28  & 35.88   & 13.99 & 15.24& 18.53 &fs1  \\
 h43 &13.210 &  $-$0.444 & 1.93 & --   &  --	 & --    &   -- & 18.62 &fs2  \\
 h44 &13.202 &  $-$0.363 & 2.19 &1.24  & 38.14   & 12.81 & 13.70& 18.51 &fs2  \\
 h45 &13.014 &  $-$0.126 & 1.87 &0.38  & --	 & 16.07 & 18.13& 19.08 &fs1  \\
 h46 &13.210 &  $-$0.142 & 2.39 &0.64  & 38.37   & 18.03 & 21.22& 20.34 &fs1  \\
 h47 &13.296 &  $-$0.069 & 2.38 &0.48  & 36.53   & 13.71 & 14.86& 19.73 &fs1  \\
 h48 &13.214 &  $-$0.320 & 2.09 &1.15  & 39.59   &  --   &   -- & 18.44 &fs2  \\
 h49 &13.420 &  $-$0.206 & 1.22 &0.43  & --	 &  --   &   -- & 18.77 &fs2   \\
 
 \hline          
\end{tabular}
\end{table*}
\subsection{ALMAGAL clumps and cores}
\label{almacores}
\begin{figure*}
   \centering
   \includegraphics[width=0.7\textwidth]{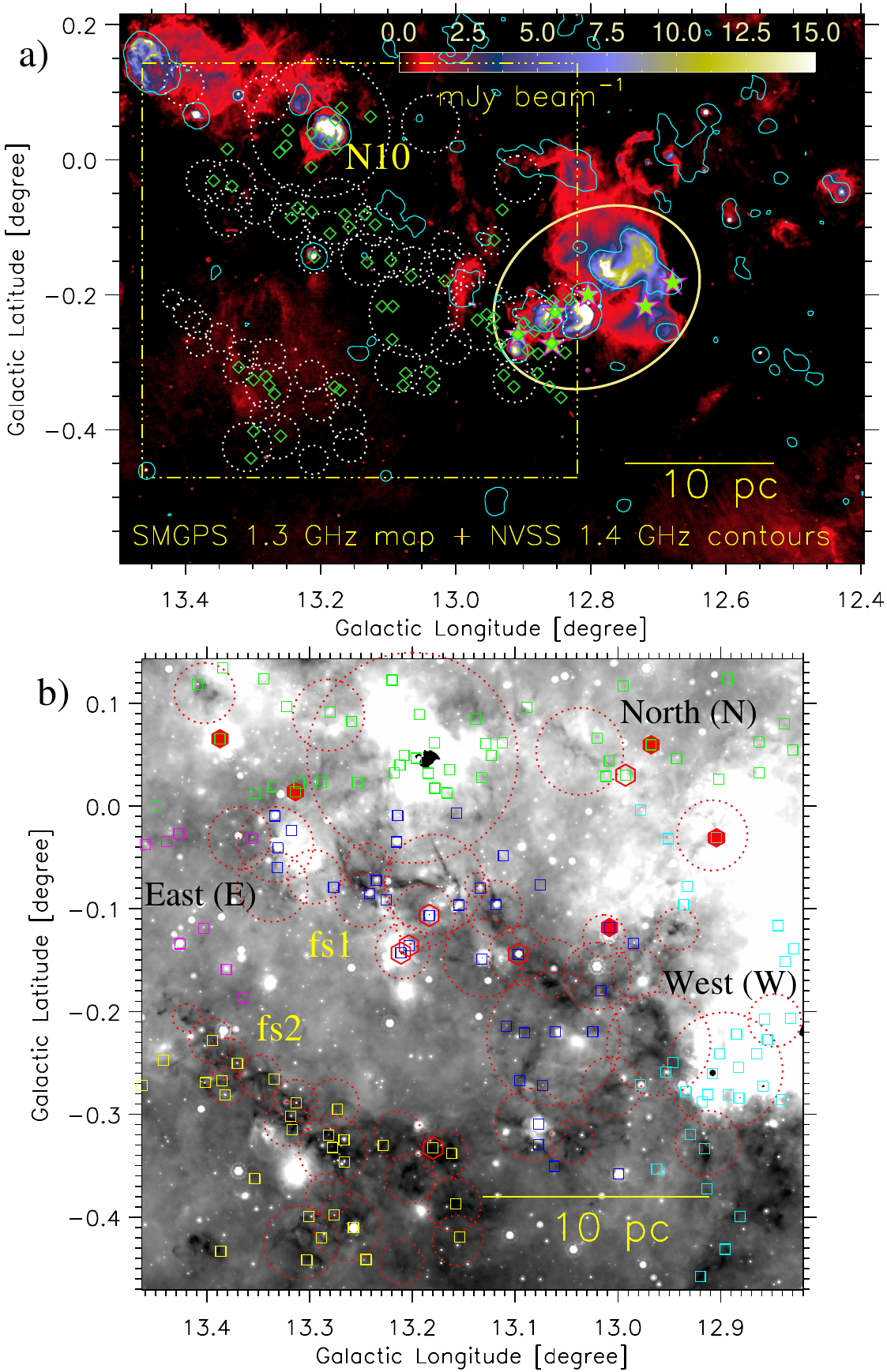}\\
 \caption{a) Overlay of the positions of HFSs (dotted circles), ATLASGAL clump locations \citep[diamonds; from][]{urquhart18}), and the NVSS 1.4 GHz continuum emission contour at 4.7 mJy 
 beam$^{-1}$ on the SMGPS 1.3 GHz continuum map. b) The panel displays the 24 $\mu$m image overlaid with the positions of HFSs (dotted circles) and Hi-GAL clumps \citep[squares; from][]{elia17}. Open and filled hexagons indicate the positions of candidate HFSs \citep[from][]{kumar20}.}
\label{fig:apx2}
\end{figure*}

\begin{figure*}
\center
\includegraphics[width=0.7\linewidth]{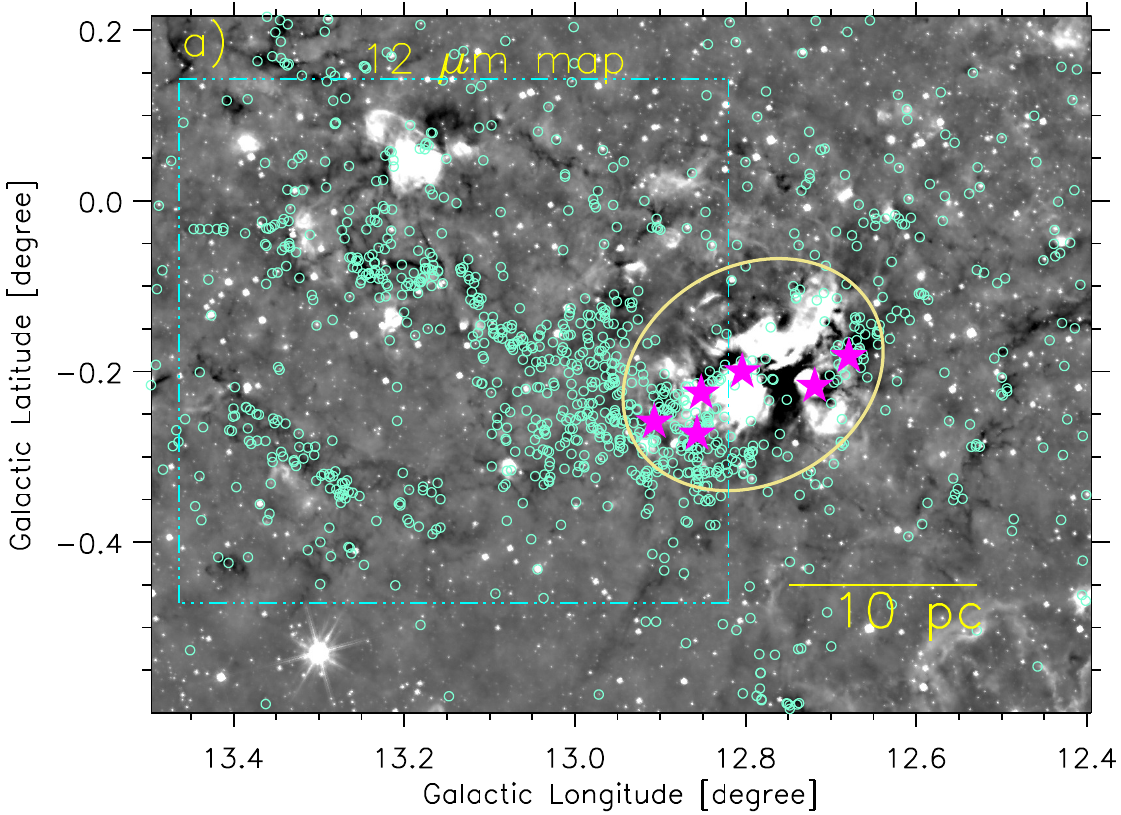}
\includegraphics[width=0.48\linewidth]{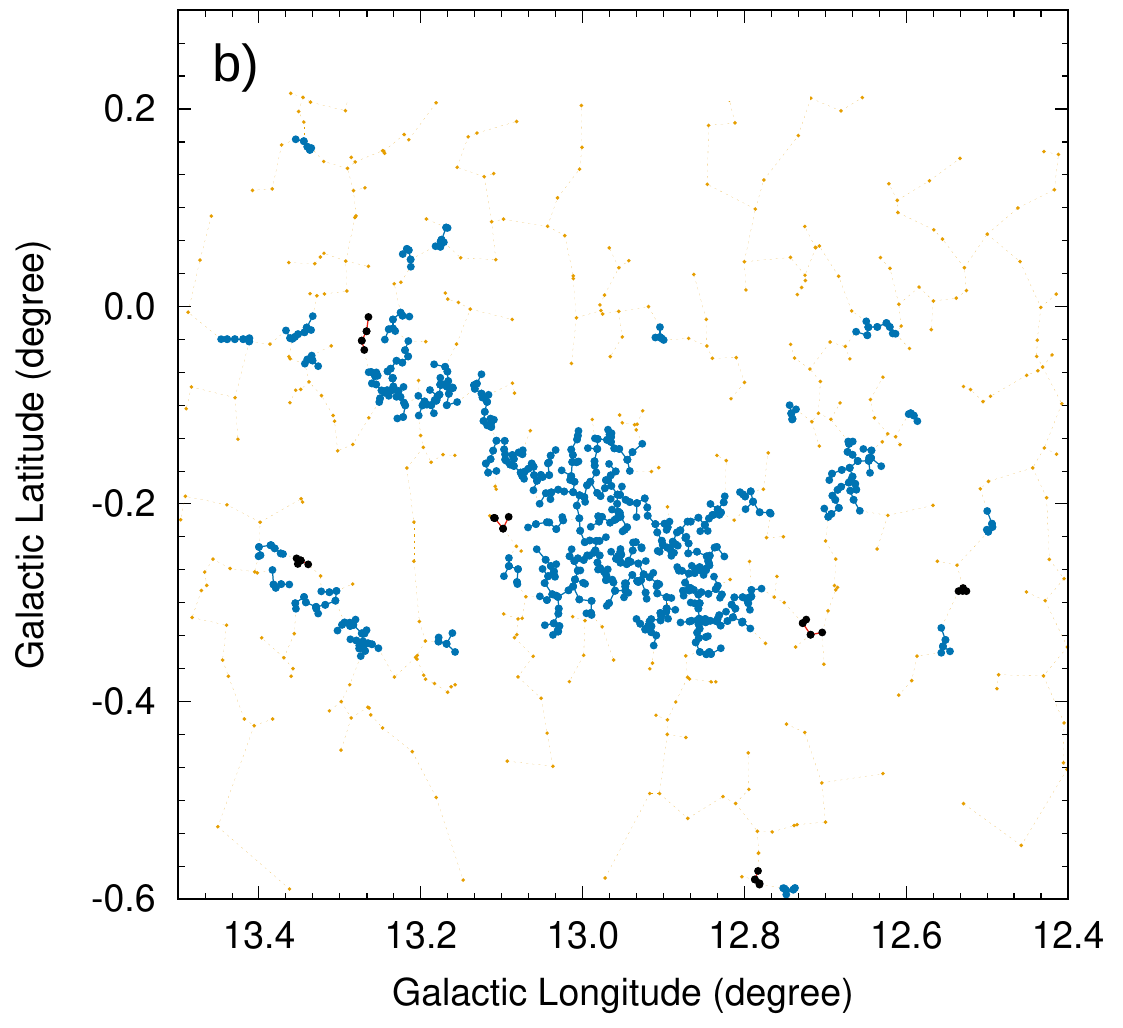}
\includegraphics[width=0.5\linewidth]{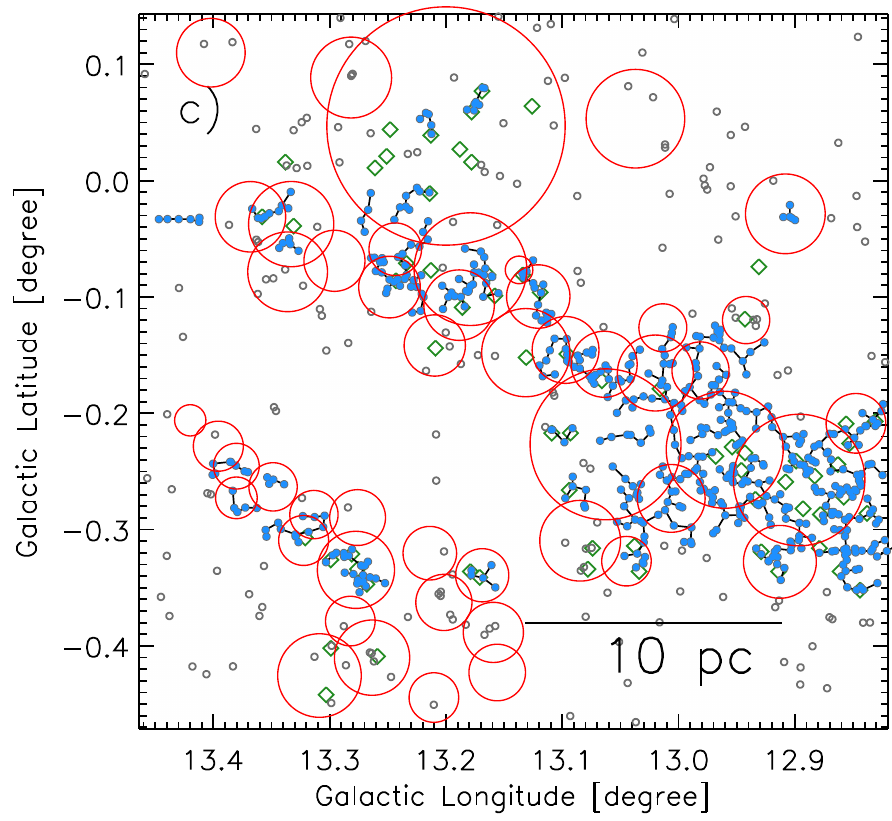}
\caption{Spatial distribution of protostars using MST approach.  
a) Overlay of protostar positions (open circles) on the 12 $\mu$m image. 
The ellipse and filled stars are the same as shown in Figure~\ref{fig1}a. 
b) MST of protostar positions in the area shown in Figure~\ref{fig1}a. 
Blue circles connected by solid lines represent MST branches smaller than the critical branch length, indicative of core-scale groupings. 
c) MST of protostar positions distributed in the area outlined by the dot-dashed box in 
Figure~\ref{fig3}a (also see Figure~\ref{fig3}b). Large red circles indicate the positions of HFSs (also see Figure~\ref{fig1}b), and diamonds mark the locations of 870 $\mu$m dust continuum clumps.}
\label{fig3}
\end{figure*}

\begin{figure*}
\centering
\includegraphics[width=0.47\textwidth]{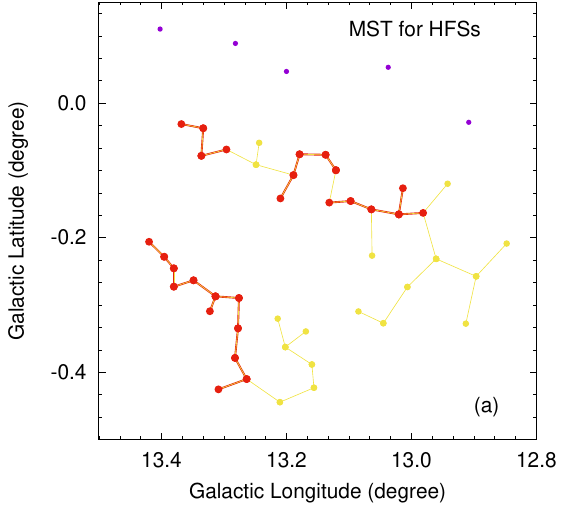}
\includegraphics[width=0.44\textwidth]{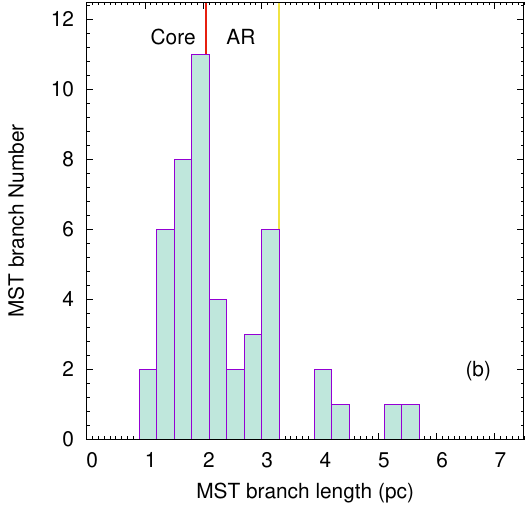}
\includegraphics[width=0.44\textwidth]{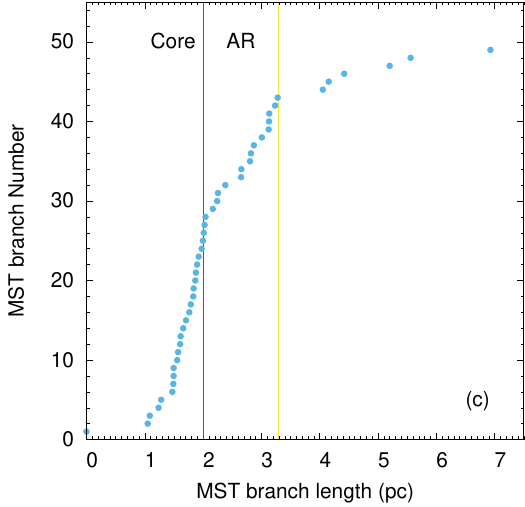}
\caption{a) MST of the identified HFSs in the area presented in Figure~\ref{fig1}b. 
Solid lines connecting red and yellow dots represent MST branches shorter than the critical lengths of 2 pc for cores and 3.3 pc for active regions (ARs), respectively (see text for more details). 
Purple dots indicate isolated HFSs that are not connected by branches shorter than the 
respective critical lengths. b) Histogram showing the distribution of MST branch lengths for the HFSs. 
The red and yellow vertical lines mark the critical branch lengths for cores (2 pc) and ARs (3.3 pc), respectively.
c) Cumulative distribution of MST branch lengths for the HFSs. 
The red and yellow vertical lines indicate the critical branch lengths for cores (2 pc) and ARs (3.3 pc), respectively.}
\label{mst}
\end{figure*}

\begin{figure*}
\center
\includegraphics[width=0.6\linewidth]{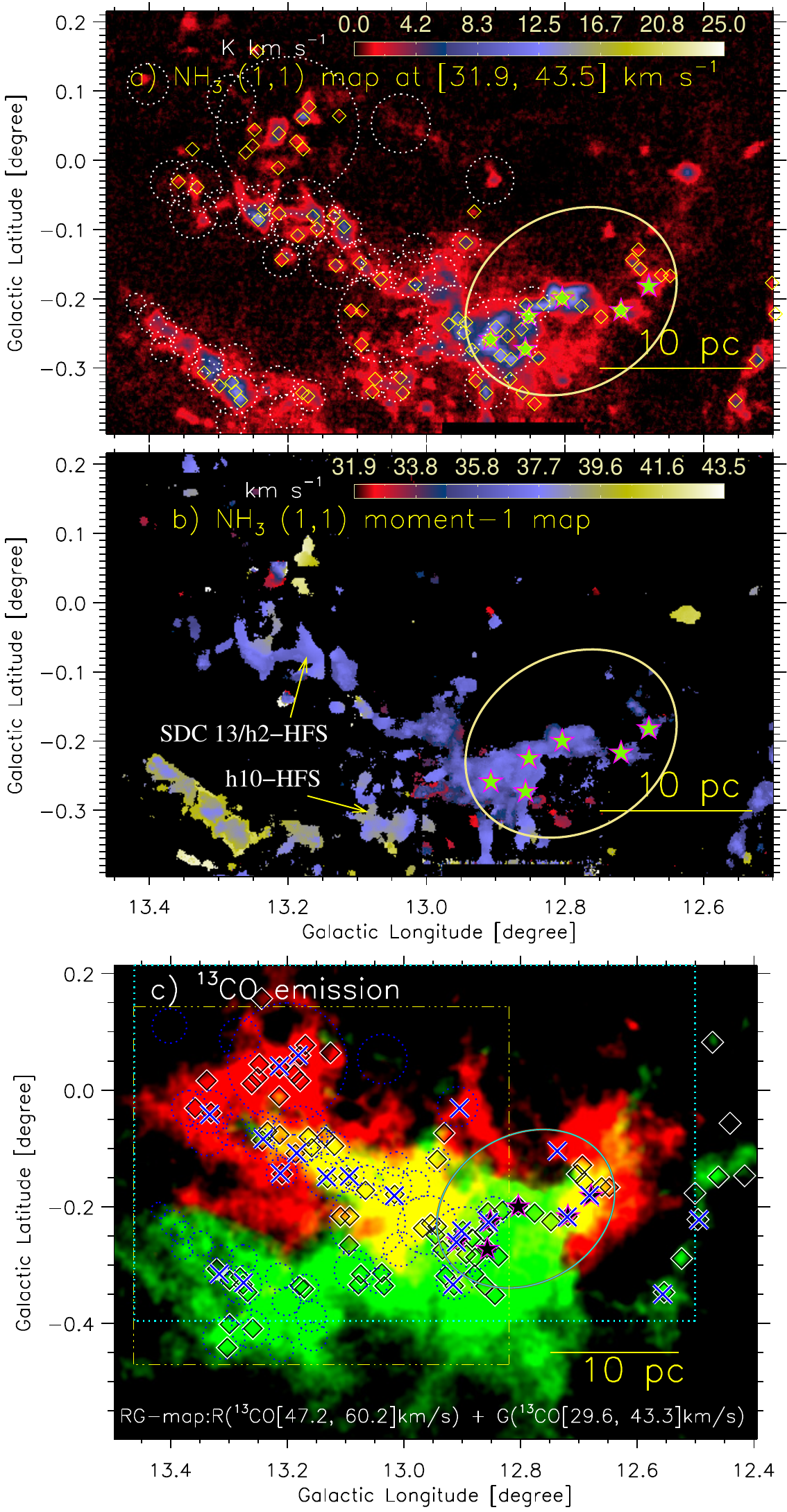}
\caption{a) Overlay of the positions of HFSs (dotted circles) and 870 $\mu$m dust continuum clumps (diamonds) on the RAMPS NH$_{3}$(1,1) moment-0 map integrated over the velocity range of [31.9, 43.5] km s$^{-1}$. 
b) RAMPS NH$_{3}$(1,1) moment-1 map of the same area. 
c) Overlay of several positions covered in the ALMAGAL 1.38 mm continuum observations \citep[from][]{coletta25} 
marked by multiplication symbols on a two color composite image, which is made with the $^{13}$CO maps at [47.2, 60.2] km s$^{-1}$ (red) and [29.6, 43,3] km s$^{-1}$ (green) \citep[from][]{dewangan20w33}. Diamonds and dotted circles are the same as in Figure~\ref{fig4}a. The dotted box indicates the area presented in Figures~\ref{fig4}a and~\ref{fig4}b, while the dot-dashed box is the same as shown in Figure~\ref{fig1}a. In each panel, the ellipse and filled stars are the same as shown in Figure~\ref{fig1}a.}
\label{fig4}
\end{figure*}

We also examined the spatial distribution of dust continuum compact sources within our target field using high-angular resolution ALMA 1.38 mm continuum data (beam size $\sim$0\rlap.{$''$}15--0\rlap.{$''$}3) from the ALMAGAL project \citep{molinari25}. 
The survey targets 1013 massive dense clumps (M $>$ 500 M$_{\odot}$, d $\leq$ 7.5 kpc) spanning evolutionary stages from IRDCs to H\,{\sc ii} regions, distributed from the near tip of the Galactic Bar to the outer Galaxy. Most recently, \citet{coletta25} released the ALMAGAL catalog of 1.38 mm compact continuum sources identified in these clumps. Among these, 21 clumps are found toward the area presented in Figure~\ref{fig:apx2}a (see multiplication symbols in Figure~\ref{fig4}a). 
However, in the target area highlighted by the dot-dashed box in Figure~\ref{fig:apx2}a, 
16 ALMAGAL clumps (number of cores; mass range), which are observed with ALMA (beam size $\sim$0\rlap.{$''$}3), 
are AG012.8535-0.2265 (11; 0.13--3.95 M$_{\odot}$), AG012.9008-0.2404 (10; 0.06--1.04 M$_{\odot}$), AG012.9048-0.0306 (14; 0.15--5.59 M$_{\odot}$), AG012.9084-0.2604 (23; 0.06--3.23 M$_{\odot}$), AG012.9156-0.3341 (8; 0.21--7.78 M$_{\odot}$), AG013.0166-0.1797 (1; 0.34 M$_{\odot}$), 
AG013.0965-0.1454 (7; 0.19--0.96 M$_{\odot}$), AG013.1319-0.1498 (13; 0.96--11.82 M$_{\odot}$), AG013.1787+0.0603 (24; 0.19--5.08 M$_{\odot}$), AG013.1845-0.1071 (7; 0.17--2.61 M$_{\odot}$), AG013.2105-0.1440 (17; 0.09--1.28 M$_{\odot}$), AG013.2123+0.0404 (4; 0.13--0.81 M$_{\odot}$), AG013.2428-0.0854 (22; 0.31--4.25 M$_{\odot}$), AG013.2775-0.3323 (3; 0.49--4.60 M$_{\odot}$), AG013.3167-0.3145 (2; 0.6--1.71 M$_{\odot}$), AG013.3313-0.0395 (3; 0.28--0.74 M$_{\odot}$). More details of the clumps/cores can be found in \citet{coletta25}. These clumps are associated with some selected HFSs examined in this study. 
The core mass distribution spans from a minimum of 0.06 M$_{\odot}$ (seen in AG012.9008-0.2404 and AG012.9084-0.2604) to a maximum of 11.82 M$_{\odot}$ (in AG013.1319-0.1498). 
The cumulative distribution of cores shows that the majority of cores have masses below 5 M$_{\odot}$. 
It indicates that the ALMAGAL clumps are primarily composed of low-mass cores. It implies that some HFSs are characterized by such low-mass core populations. 

MST analysis was performed for the continuum cores associated with 11 clumps, corresponding to ten HFSs, each containing more than five cores (not shown). We examined the histogram of the MST branch lengths derived from these cores (see Figure~\ref{figuu5}  and also Appendix~\ref{susec:gsp}). The resulting median nearest-neighbor separation across the entire sample is estimated to be $\sim$0.03 pc.
\section{Discussion}
\label{sec:disc}
Small-scale/compact HFSs (extent $<$ 3 pc) were commonly identified in previous high-resolution Atacama Large Millimeter/submillimeter Array (ALMA) studies targeting young clusters and massive star-forming regions \citep[e.g.,][]{zhou22}. Key models explaining star formation in HFSs include the global non-isotropic collapse (GNIC) scenario \citep{Tige+2017,Motte+2018}, the Filaments to Clusters (F2C) model \citep{kumar20}, and the inertial inflow model \citep{padoan20}. The GNIC scenario incorporates elements of both the competitive accretion (CA) \citep{bonnell01,bonnell04} and global hierarchical collapse (GHC) models \citep{semadeni09,semadeni17,semadeni19}. 
More details of different existing scenarios related to HFSs can be found in \citet{bhadari25}.

The present work focuses on 45 HFSs seen in absorption with a median size of $\sim$2.4 pc (see Section~\ref{s1sec:d}).
This section discusses the implications of the observed results in addressing some of the key questions outlined in Section~\ref{sec:intro}.  

From Section~\ref{sec:msta}, it is found that protostars are strongly associated with HFSs, with protostellar core separations of $\leq$ 0.7 pc. These HFSs fall into two main groupings that span 10--15 pc, with separations between them ranging from 1 to 3.3 pc. About 65\% of the HFSs form tight clusters (separations $<$ 2 pc), displaying a rich population of small-scale structures. In the selected target area, we find 16 ALMAGAL clumps observed with ALMA (see Section~\ref{almacores} and multiplication symbols within the dot-dashed box in Figure~\ref{fig4}c).
Among them, 11 clumps associated with ten HFSs are resolved into more than five cores each, with the number of detected cores ranging from $\sim$7 to 24. MST analysis of these cores reveals a median nearest-neighbor separation of $\sim$0.03 pc (see Section~\ref{almacores} for more details).  
This core spacing likely represents a characteristic fragmentation scale within dense clumps, 
potentially governed by thermal Jeans instability or influenced by additional factors such as turbulence and magnetic fields. The large protostellar spacing ($\sim$0.7 pc) compared to the Jeans length ($\sim$0.08 pc for temperature $\sim$18 K 
and density $\sim$10$^{5}$ cm$^{-3}$) suggests that thermal fragmentation alone is insufficient to account for the observed 
larger-scale protostellar distribution, though the smaller core spacing remains consistent with thermally driven fragmentation.
Recently, \citet{xu24} reported that cores in the evolved sample exhibit a characteristic separation of $\sim$0.03 pc, compared to $\sim$0.1 pc in the earlier evolutionary stage. In the early phase, such separations are consistent with fragmentation driven by thermal Jeans instability. However, as the HFS contracts, the typical core spacing decreases. 
The smaller separations observed in W33 appear consistent with this evolutionary picture.

In each IRDC, HFSs are predominantly located at the junctions of substructures, where they are associated with dense protostellar clusters 
and show no detectable radio continuum emission. Furthermore, NH$_{3}$ clumps distributed toward these IRDCs meet the threshold required to potentially form massive stars (see Appendix~\ref{susec:gsp} for more details). 

Studying extended H\,{\sc ii} regions around high-luminosity sources in central hubs provides an effective way to trace the evolutionary stages of HFSs. At early stages, hubs are compact and dense with little ionization, while at later stages they become more extended and shaped by stellar feedback, including expanding ionized regions.
This framework offers insights into the evolution of HFSs and the role of massive-star feedback in shaping them \citep[e.g.,][]{Motte+2018,dewangan25s}.
High-resolution JWST and ALMA observations have uncovered a small-scale HFS (G11P1-HFS; extent $<$0.6 pc) in its nascent phase \citep{dewangan24,bhadari25}, while \citet{dewangan25s} reported that the Mon R2 HFS evolved from an early to a more advanced phase through the combined effects of gas accretion and massive-star feedback.  In the W33 complex, the study of the physical properties of HFSs suggest that those in fs1 are relatively more evolved and dynamically active, whereas those in fs2 remain cooler, smaller, and less evolved, consistent with an earlier evolutionary stage (see Section~\ref{s1sec:d}). However, no significant radio continuum emission is detected toward the HFSs in either fs1 or fs2. In contrast, the North and West regions host fewer but warmer and more evolved HFSs, consistent with more advanced star formation and the presence of extended ionized emission (see Section~\ref{sec:clump}). Similar findings are also derived using the Hi-GAL clumps (see Section~\ref{s2sec:dx}).

The presence of a large population of HFSs within a single star-forming complex supports a scenario of efficient and clustered star formation capable of forming massive stars. 
These findings favour a picture in which cloud or filament fragments collapse into clumps, each hosting a HFS that drives active star formation. Such compact or small-scale HFSs appear to be a common structural element in both cluster-forming and massive star-forming regions, playing a pivotal role in the early stages of stellar assembly. This allows us to suggest that, prior to the onset of MSF or the development of H\,{\sc ii} regions in the W33 complex, the region may have contained numerous compact or small-scale HFSs. These HFSs were embedded toward overlapping filamentary substructures, where massive stars formed and evolved.  

The ubiquitous association of star formation with HFSs is highlighted in \citet{myers09}, who presented various morphologies of HFSs across different nearby star-forming regions, including those forming massive stars (see Figures 1--15 in their work).
Similar to the W33 complex, the GLIMPSE/MIPSGAL image of G345.00$-$0.22 also shows multiple HFSs \citep[see Figure 15 in][]{myers09}. Nevertheless, the underlying processes responsible for the origin of such complex systems hosting numerous HFSs remain poorly understood. 

W33 is situated at the intersection of the Scutum and Norma spiral arms (see the near sides of these arms highlighted in the NH$_3$(1,1) longitude-velocity diagram in Figure~\ref{fig:apx4}). Such spiral arm junctions are typically known to host enhanced gas densities, experience stronger dynamical perturbations, and support large-scale inflow motions, creating favorable conditions for molecular cloud collisions and converging gas flows \citep{dobbs14}. 
Such events are facilitated by arm-induced gas compression and material funneled by spiral streaming motions. 
As a result, increased cloud-cloud interactions promote gravitational collapse, leading to the formation of new stars \citep{fukui21b}.
Earlier studies, based on the kinematics and spatial distribution of CO molecular gas, suggested 
an onset of a cloud-cloud collision (CCC) scenario toward the W33 complex \citep[e.g.,][]{dewangan20w33,zhou23}. 
Using the $^{13}$CO emission, the spatial distribution of two cloud components at [29.6, 43,3] and [47.2, 60.2] km s$^{-1}$ \citep[from][]{dewangan20w33} is presented in Figure~\ref{fig4}c. 

The hydrodynamic simulations by \citet{balfour15} demonstrated that low-velocity cloud collisions under subsonic turbulence form radially converging filaments, while higher-velocity collisions produce spider web-like structures that give rise to numerous HFSs (see Figure 3 in their work). However, under supersonic turbulence or pre-collision inhomogeneous density structures, 
this morphological distinction becomes less evident \citep{Balfour_2017MNRAS}. 
The W33 complex, with a velocity dispersion exceeding 2 km s$^{-1}$, exhibits signs of supersonic turbulence \citep[see Figure 1e in][]{zhou23}.
Magneto-hydrodynamic simulations further reveal that, in CCCs involving supersonic turbulence, the turbulence creates inhomogeneous density structures prior to the collision, and shock compression during the interaction forms filaments. Magnetic fields shape these structures by opposing gas motions perpendicular to the field lines \citep[e.g.,][]{inoue13,inoue18,fukui21b,Maity_2024}. 
As shown in Figure~1 of \citet{fukui21b}, the collision event can give rise to multiple HFSs. 
Furthermore, filaments may break into smaller-scale sub-filaments under the combined effects of 
residual turbulence and self-gravity \citep{Tafalla_2015A&A}. 

Overall, the observed morphology and spatial organization in W33 are likely arises from collisions of supersonic turbulent gas flows, together with self-gravity and residual turbulence. 
It appears that several physical processes as mentioned earlier are underway. 
\section{Summary and Conclusion}
\label{sec:conc}
We carried out a multi-wavelength, multi-scale study of the W33 complex, which is located at the junction of the Scutum and Norma spiral arms.
The focus of this study is to understand the physical processes driving intense star formation in the complex. 
Our main results are as follows:\\
$\bullet$ In the W33 complex, 45 compact HFSs (median size $\sim$2.4 pc) in IRDCs are investigated using {\it Spitzer} 8 and 24 $\mu$m, and unWISE 12 $\mu$m images (resolution $\sim$2$''$--6$''$). These HFSs are identified by the presence of three or more filaments converging onto a central hub, detected as absorption features toward IRDCs. This represents the first report of such a high concentration of small-scale HFSs (extent $\sim$2--3 pc) within a single star-forming complex.\\ 
$\bullet$ The HFSs are located at the intersection of elongated infrared dark substructres, which are associated with clusters of protostars, and lack radio continuum emission. \\
$\bullet$ NH$_{3}$ emission from the RAMPS survey (resolution $\sim$34\rlap.{$''$}7) is detected toward both the IRDCs (including HFSs) and the central W33 region. \\
$\bullet$ Based on their physical properties and the absence of radio continuum emission, the majority of the selected compact HFSs appear to be in an earlier evolutionary stage.\\
$\bullet$ MST analysis shows that protostars are strongly clustered in HFSs, with separations of $\leq$0.7 pc. The HFSs form two main groupings spanning 10--15 pc, with member separations of 1--3.3 pc. 
About $\sim$65\% of them are tightly clustered within $<$2 pc, exhibiting rich substructures.\\
$\bullet$ Applying MST analysis to the ALMAGAL 1.38 mm continuum cores, predominantly low-mass, distributed across ten HFSs (each containing more than five cores) reveals a median core spacing of $\sim$0.03 pc.\\
$\bullet$ The protostellar spacing ($\sim$0.7 pc) exceeds the thermal Jeans 
length ($\sim$0.08 pc for temperature $\sim$18 K, density $\sim$10$^{5}$ cm$^{-3}$), while the core 
spacing is smaller than the Jeans length. This indicates that thermal fragmentation plays 
a role at core scales but cannot fully explain the large-scale protostellar distribution.\\

Overall, our findings support a scenario in which cloud/filament fragments form compact HFSs that promote efficient, clustered star formation, frequently leading to MSF. 
The global structure of W33 appears to be governed by supersonic gas collisions, self-gravity, and residual turbulence. 

We thank the anonymous referees for useful comments. The research work at Physical Research Laboratory is funded by the Department of Space, Government of India. RKY gratefully acknowledges the support from the Fundamental Fund of Thailand Science Research and Innovation (TSRI, Confirmation No. FFB680072/0269) through the National Astronomical Research Institute of Thailand (Public Organization). We thank Dr. M. S. N. Kumar for the initial discussion.  This work was performed in part at the Jet Propulsion Laboratory, California Institute of Technology, under contract with the National Aeronautics and Space Administration (80NM0018D0004).  
This publication makes use of molecular line data from the Radio Ammonia Mid-Plane Survey (RAMPS). 
RAMPS is supported by the National Science Foundation under grant AST-1616635.
%
\vspace{5mm}
\facilities{Spitzer, unWISE, Herschel, APEX/ATLASGAL, GTB/RAMPS}
%
\appendix
\restartappendixnumbering
\section{Hub-filament systems and NH$_{3}$ clumps toward IRDCs}
\label{susec:gsp}
\begin{figure*}
   \centering
  \includegraphics[width=1\textwidth]{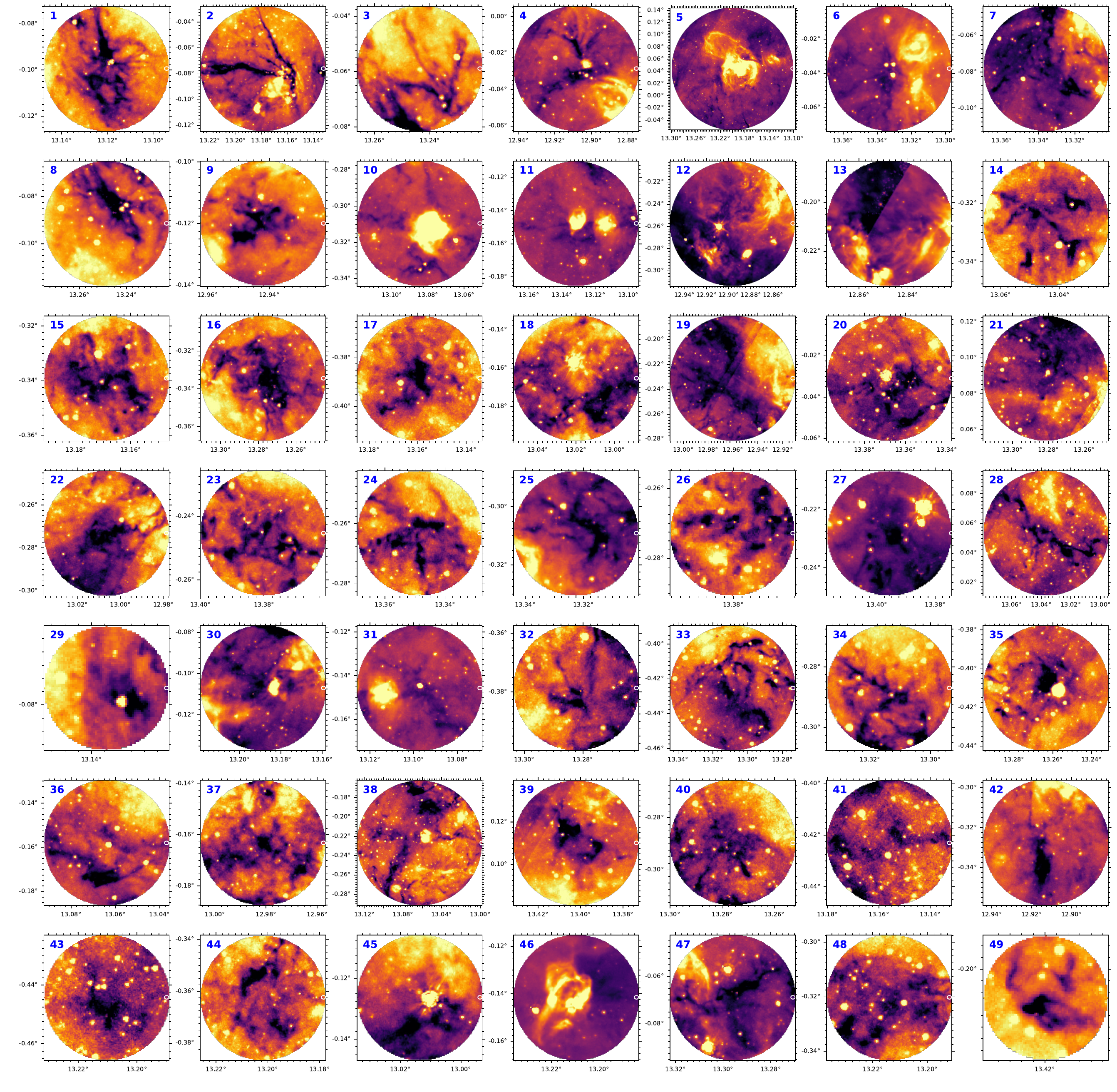}\\
\caption{A zoomed-in view of each HFS is shown using the 8 $\mu$m image (see Figure~\ref{fig:apx2}a and also Table~\ref{tab1}). 
The area presented for each HFS corresponds to its respective size as listed in Table~\ref{tab1}.} 
\label{fig:apx1gg}
\end{figure*}

MIR images ($\sim$2$''$--6$''$) obtained from space at wavelengths of 8, 12, and 24 $\mu$m were examined toward the large-scale 
physical environment of the W33 complex. These images reveal intertwined substructures along with HFSs appearing in absorption. Figure~\ref{fig:apx1gg} presents zoomed-in views of individual HFSs using the 8 $\mu$m image, where the displayed area for each case corresponds to its respective size listed in Table~\ref{tab1}.

Previously, a large-scale HFS centered around H\,{\sc ii} regions in W33 has also been reported 
using $^{12}$CO/$^{13}$CO/C$^{18}$O line data with relatively coarse beam sizes \citep{liu21,zhou23}. 
The spatial distribution of the $^{13}$CO maps at [47.2, 60.2] km s$^{-1}$ (red) and [29.6, 43,3] km s$^{-1}$ (green) \citep[from][]{dewangan20w33} is presented in 
Figure~\ref{fig4}c. To examine the NH$_{3}$(1,1) emission, the NH$_{3}$(1,1) line data from the RAMPS survey are used in this paper (see the dotted box in Figure~\ref{fig4}c). 
The survey maps molecular line emission along the Galactic midplane in the first quadrant of the Galaxy, covering the range {\it l} = $10^\circ \leq \ell \leq 40^\circ$, $|{\it b}| \leq 0.4^\circ$. In the direction of the selected target area hosting W33, the NH$_{3}$(1,1) emission appears more extended and intense compared to other areas surveyed \citep{hogge18}. 

\begin{figure*}
   \centering
  \includegraphics[width=10cm]{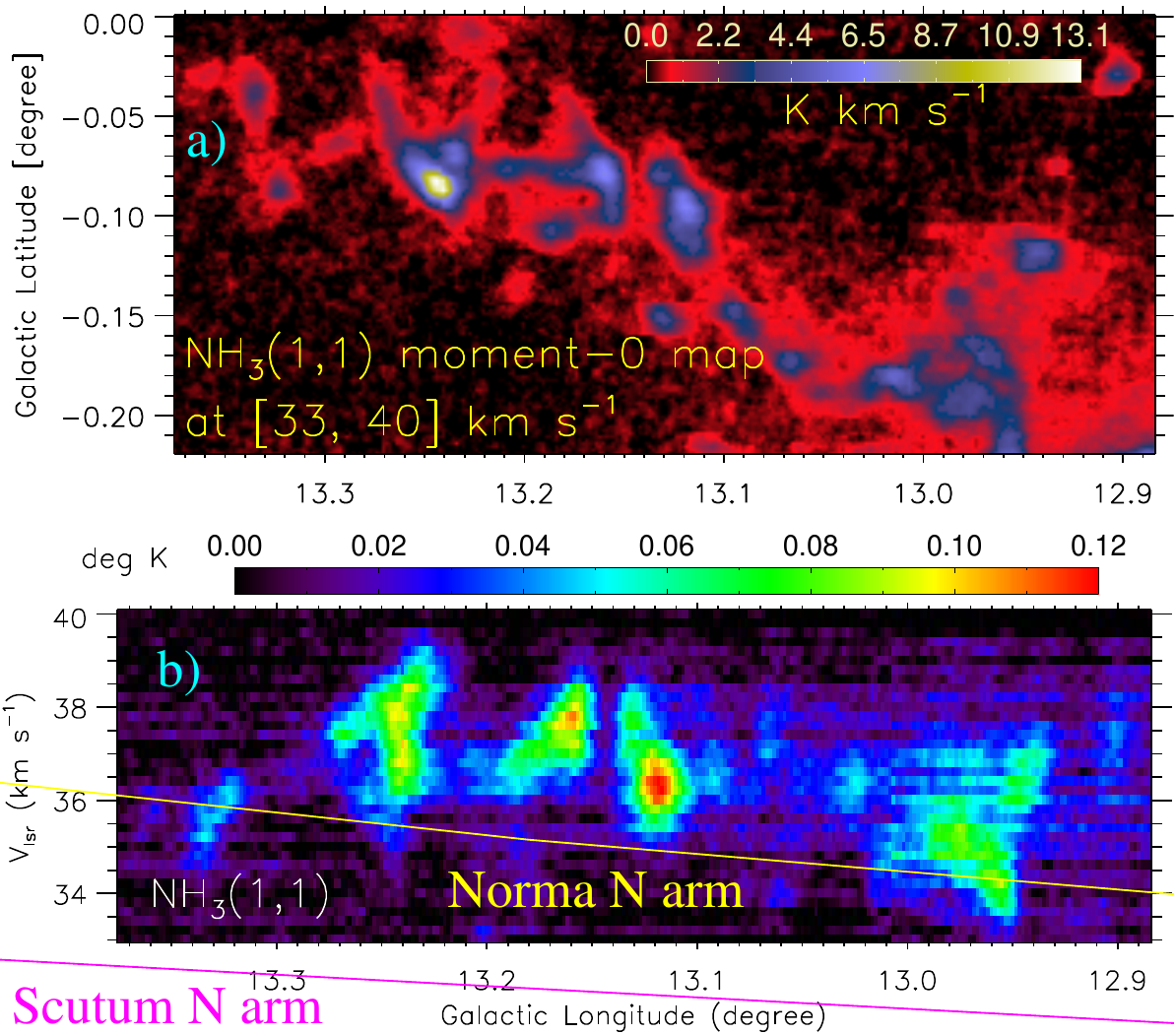}\\
  \includegraphics[width=10cm]{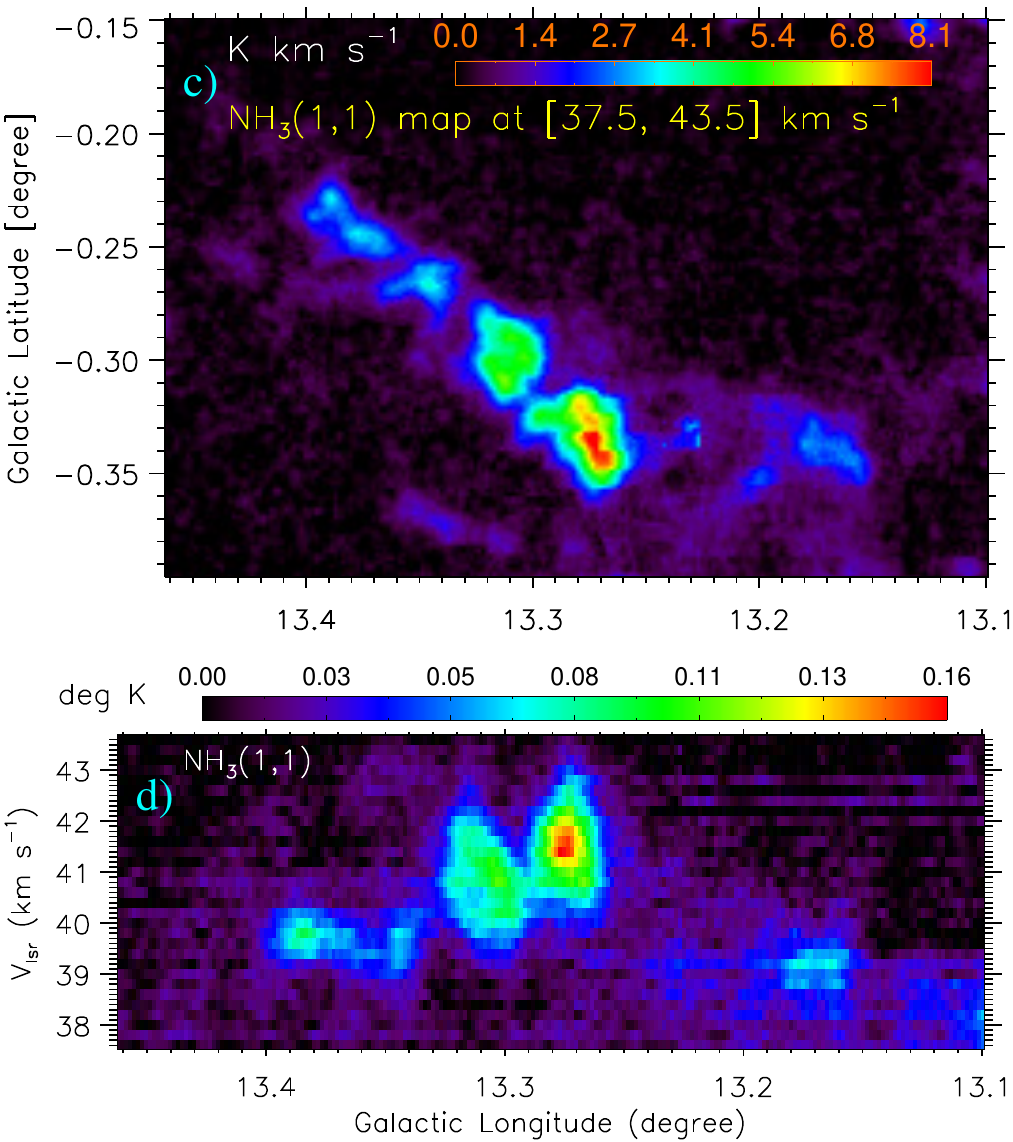}\\
\caption{a) NH$_3$(1,1) moment-0 map at [33, 40] km s$^{-1}$ of the area hosting the filament fs1 (see Figure~\ref{fig1}b). b) NH$_3$(1,1) longitude-velocity diagram for Galactic latitude integration range of [$-$0$^\circ$.219, 0$^\circ$.0] (see Figure~\ref{fig:apx4}a). The near sides of the Scutum and Norma arms \citep[from][]{reid16} are also highlighted on the NH$_3$(1,1) longitude-velocity diagram. c) NH$_3$(1,1) moment-0 map at [37.5, 43.5] km s$^{-1}$ of the area containing the filament fs2 (see Figure~\ref{fig1}b). 
d) NH$_3$(1,1) longitude-velocity diagram for Galactic latitude integration range of  [$-$0$^\circ$.396, $-$0$^\circ$.15] (see Figure~\ref{fig:apx4}c).}
\label{fig:apx4}
\end{figure*}

In the direction of fs1, NH$_{3}$(1,1) emission is studied in a velocity range of [33, 40] km s$^{-1}$ (see Figure~\ref{fig:apx4}a), while for fs2, the emission is analyzed in a velocity range of [37.5, 43.5] km s$^{-1}$ (see Figure~\ref{fig:apx4}c). 
To explore the velocity structures toward these filaments, position-velocity diagrams are produced (Figures~\ref{fig:apx4}b and~\ref{fig:apx4}d). In each filament, several molecular clumps with distinct velocity spreads are found. The velocity differences observed between fs1 and fs2 may indicate variations in their dynamical 
states or their relative positions along the line of sight. 
The near sides of the Scutum and Norma arms, based on the spiral structure outlined by \citet{reid16}, are also marked in In Figure~\ref{fig:apx4}b. 

\begin{figure}
\center
\includegraphics[width=0.8\linewidth]{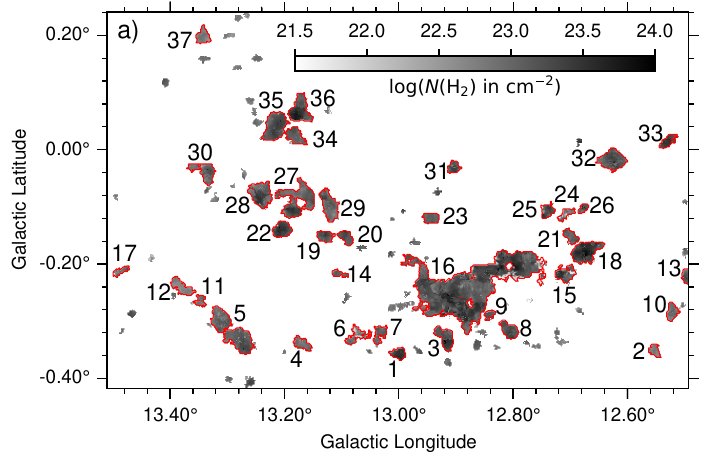}
\includegraphics[width=0.58\linewidth]{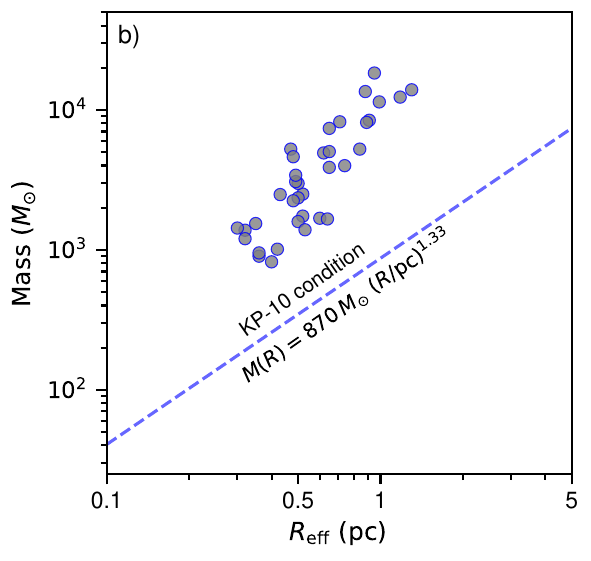}
\caption{a) Overlay of clumps identified with the {\texttt{astrodendro}} algorithm (shown as red contours) 
on the H$_2$ column density ($N$(H$_2$)) map, derived from the NH$_3$ column density data obtained from the RAMPS survey. Each NH$_3$ clump is labeled in the panel (see Table~\ref{tab2} in the Appendix for details). b) Mass-radius ($M_{\mathrm{clump}}$--$R_{\mathrm{eff}}$) plot for the NH$_{3}$ clumps. 
The blue dashed line marks the empirical threshold for MSF proposed by \citet{Kauffmann2010}.}
\label{fig5}
\end{figure}

The survey also offers derived data products, including the NH$_{3}$ column density map ($N$(NH$_3$)). 
Using the abundance ratio, $\frac{N(\mathrm{NH}_3)}{N(\mathrm{H}_2)} \sim 2 \times 10^{-8}$ \citep{Pineda2022AJ}, 
the existing RAMPS $N$(NH$_3$) map was converted into an $N$(H$_2$) column density map. 
Figure~\ref{fig5}a presents the $N$(H$_2$) map of the selected target area, which is derived using the existing $N$(NH$_3$) map collected from the RAMPS survey \citep[see][for more details]{hogge18}. To identify clumps from the $N$(H$_2$) map of the target area, the Python-based tool \href{https://dendrograms.readthedocs.io/en/stable/index.html}{\texttt{astrodendro}} \citep{Rosolowsky_2008ApJ} was utilized. The clumps identified by {\texttt{astrodendro}} are highlighted in Figure~\ref{fig5}a. Some of these clumps also exhibit hierarchical substructures; however, in this study, we focus only on the major structures. 
The locations of these clumps in Galactic coordinates and effective radii ($R_{\mathrm{eff}}$) are listed in Table~\ref{tab2}. 
Clump radii were computed using the relation $R = \sqrt{Area/\pi}$, corrected for the beam size of the $N$(H$_2$) map, and then converted into physical units to obtain $R_{\mathrm{eff}}$.
The masses ($M_\mathrm{clump}$) of the clumps indicated in Figure~\ref{fig5}a were determined using the following equation: \begin{equation} 
M_{\mathrm{clump}} = \mu_{\mathrm{H}_2}\, m_{\mathrm{H}}\, {A}_{\mathrm{pixel}}\, \sum N(\mathrm{H}_2), 
\end{equation} 
where $\mu_{\mathrm{H}_2}$ is the mean molecular weight \citep[adopted to be 2.8;][]{Kauffmann_2008}, $m_{\mathrm{H}}$ is the mass of a hydrogen atom, $\sum N$(H$_2$) is the total H$_2$ column density, and $A_{\mathrm{pixel}}$ is the area subtended by a single pixel. 

Assuming spherical symmetry, the expressions used to compute the average volume density ($n$) and column density ($N(\mathrm{H}_2)$) of the clumps are given by $n = \frac{3M_\mathrm{clump}}{4\pi \mu m_{\mathrm{H}} R_{\mathrm{eff}}^3}$ and $N(\mathrm{H}_2) = \frac{M_\mathrm{clump}}{\pi \mu m_{\mathrm{H}} R_{\mathrm{eff}}^2}$, respectively. 
The mass, average density, and average column density values for the clumps are listed in Table~\ref{tab2}.
It is important to note that the NH$_3$ abundance ratio is known to decrease above a column density of $N$(H$_2$) = 2.575 $\times 10^{22}$ cm$^{-2}$ \citep{Pineda2022AJ}. As a result, the derived values of mass and average column density represent lower limits. Figure~\ref{fig5}b presents the $M_{\mathrm{clump}}$--$R_{\mathrm{eff}}$ plot for the identified clumps. All molecular clumps are found to satisfy the Kauffmann \& Pillai condition for MSF \citep[][]{Kauffmann2010}. Note that the selected HFSs are found toward the molecular clumps. 

\begin{figure}
\center
\includegraphics[width=\linewidth]{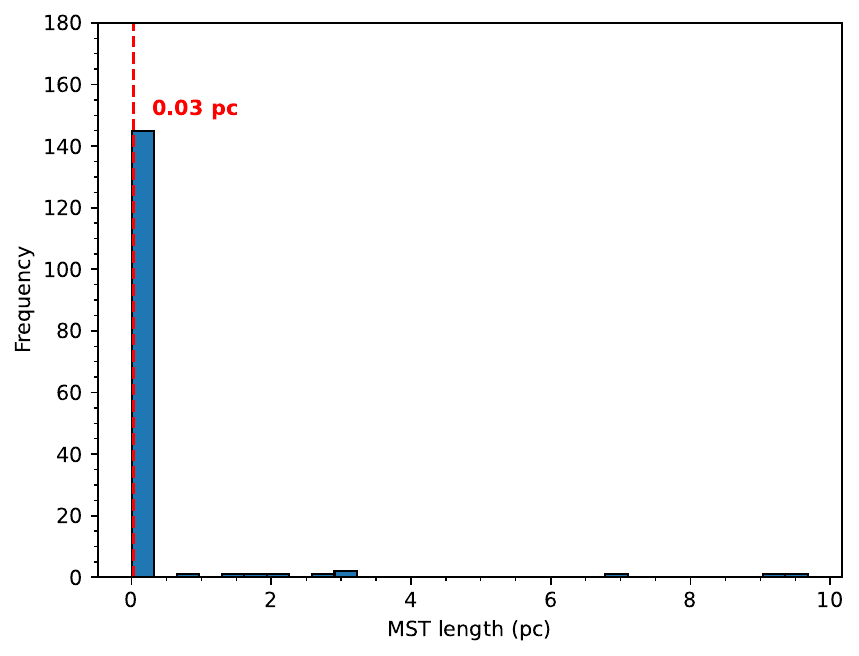}
\caption{Histogram displaying the distribution of MST branch lengths for the ALMAGAL 1.38 mm continuum cores associated with ten HFSs (each containing more than five cores).} 
\label{figuu5}
\end{figure}

The ALMAGAL cores associated with 11 clumps (corresponding to ten HFSs, each containing more than five compact cores) were analyzed using the MST method. The histogram of MST branch lengths is presented in Figure~\ref{figuu5}. From this distribution, a median nearest-neighbor core separation is estimated to be $\sim$0.03 pc, as marked in the figure.
\begin{table*}
\small
\setlength{\tabcolsep}{0.05in}
\centering
\caption{Table lists physical properties of molecular clumps derived from the H$_2$ column density ($N$(H$_2$)) map, which is estimated from the NH$_3$ column density data obtained from the RAMPS survey (see Figure~\ref{fig5}a and appendix~\ref{susec:gsp}). 
The listed parameters include the clump ID, coordinates, effective radius ($R_{\mathrm{eff}}$), mass, average volume density, and average column density ($N$(H$_2$)).} 
\label{tab2}
\begin{tabular}{lcccccccccccccc}
\hline  
ID & $l$ &  $b$  &$R_{\rm {eff}}$&      Mass &  Density&  $N$(H$_2$)  \\
& (degree) &  (degree) &   (pc) &    ($\times 10^3 M_{\odot}$) &  ($\times 10^5$ cm$^{-3}$) &  log(cm$^{-2}$) \\
\hline
   1 & 13.000 & $-$0.357 &   0.47 &    5.24 &	1.69 &   23.52 \\
   2 & 12.553 & $-$0.351 &   0.40 &    0.82 &	0.44 &   22.86 \\
   3 & 12.918 & $-$0.331 &   0.71 &    8.21 &	0.80 &   23.37 \\
   4 & 13.169 & $-$0.339 &   0.52 &    1.74 &	0.43 &   22.97 \\
   5 & 13.289 & $-$0.322 &   1.30 &   13.90 &	0.22 &   23.07 \\
   6 & 13.073 & $-$0.326 &   0.60 &    1.68 &	0.27 &   22.82 \\
   7 & 13.033 & $-$0.322 &   0.52 &    2.50 &	0.62 &   23.12 \\
   8 & 12.807 & $-$0.316 &   0.62 &    4.92 &	0.72 &   23.26 \\
   9 & 12.842 & $-$0.289 &   0.32 &    1.38 &	1.51 &   23.29 \\
  10 & 12.522 & $-$0.283 &   0.50 &    1.59 &	0.43 &   22.95 \\
  11 & 13.347 & $-$0.264 &   0.36 &    0.90 &	0.66 &   22.99 \\
  12 & 13.380 & $-$0.240 &   0.64 &    1.66 &	0.22 &   22.77 \\
  13 & 12.499 & $-$0.222 &   0.35 &    1.54 &	1.23 &   23.25 \\
  14 & 13.104 & $-$0.217 &   0.32 &    1.20 &	1.22 &   23.21 \\
  15 & 12.713 & $-$0.217 &   0.65 &    5.03 &	0.63 &   23.23 \\
  16 & 12.874 & $-$0.239 &   3.01 &  105.87 &	0.13 &   23.22 \\
  17 & 13.485 & $-$0.212 &   0.36 &    0.95 &	0.69 &   23.01 \\
  18 & 12.674 & $-$0.178 &   0.95 &   18.32 &	0.74 &   23.46 \\
  19 & 13.130 & $-$0.151 &   0.50 &    2.98 &	0.80 &   23.22 \\
  20 & 13.093 & $-$0.152 &   0.49 &    3.07 &	0.90 &   23.26 \\
  21 & 12.699 & $-$0.152 &   0.50 &    2.35 &	0.67 &   23.13 \\
  22 & 13.205 & $-$0.140 &   0.65 &    7.37 &	0.92 &   23.39 \\
  23 & 12.944 & $-$0.120 &   0.48 &    2.24 &	0.68 &   23.13 \\
  24 & 12.708 & $-$0.113 &   0.42 &    1.01 &	0.48 &   22.91 \\
  25 & 12.740 & $-$0.108 &   0.49 &    3.41 &	1.01 &   23.31 \\
  26 & 12.677 & $-$0.102 &   0.30 &    1.43 &	1.87 &   23.36 \\
  27 & 13.178 & $-$0.092 &   1.18 &   12.33 &	0.26 &   23.10 \\
  28 & 13.242 & $-$0.080 &   0.91 &    8.42 &	0.38 &   23.16 \\
  29 & 13.123 & $-$0.095 &   0.84 &    5.24 &	0.30 &   23.02 \\
  30 & 13.338 & $-$0.039 &   0.74 &    3.99 &	0.34 &   23.01 \\
  31 & 12.904 & $-$0.032 &   0.43 &    2.48 &	1.06 &   23.28 \\
  32 & 12.626 & $-$0.018 &   0.89 &    8.13 &	0.40 &   23.17 \\
  33 & 12.530 &    0.014 &   0.48 &    4.62 &	1.42 &   23.45 \\
  34 & 13.182 &    0.025 &   0.65 &    3.88 &	0.49 &   23.11 \\
  35 & 13.215 &    0.040 &   0.99 &   11.39 &	0.41 &   23.22 \\
  36 & 13.176 &    0.065 &   0.88 &   13.51 &	0.69 &   23.40 \\
  37 & 13.343 &    0.198 &   0.53 &    1.39 &	0.32 &   22.85 \\
\hline          
\end{tabular}
\end{table*}

\bibliography{reference}{}
\bibliographystyle{aasjournal}
\end{document}